\documentclass[a4paper,11pt]{article}
\usepackage{jheppub} 
\usepackage{hyperref}
\usepackage{lineno}
\usepackage{amsmath}
\allowdisplaybreaks
\usepackage{indentfirst}
\usepackage{slashed}
\usepackage{subcaption}

\title{\boldmath Semi-analytical two-loop QCD corrections to $e^+e^-\to J/\psi+\chi_{cJ}$ at B factories}

\author[a]{Hao-Yang Liu,}
\author[a]{Cong Li,}
\author[a]{Wen-Long Sang}

\affiliation[a]{School of Physical Science and Technology, Southwest University, Chongqing 400700, China}

\emailAdd{tauceti@email.swu.edu.cn}
\emailAdd{lc312321@163.com}
\emailAdd{wlsang@swu.edu.cn~(corresponding author)}

\abstract{
In this work, we compute the next-to-next-to-leading-order (NNLO) QCD corrections to the process $e^+e^-\to J/\psi+\chi_{cJ}$ at B factories within the NRQCD factorization framework. The helicity amplitudes are obtained via asymptotic expansions around $r=0$ and $r=1$, with $r=16m_c^2/s$. Our asymptotic expressions reproduce the exact numerical results with high accuracy across the kinematic range $0\le r \le 1$, achieving a relative error below $10^{-5}$, which is sufficient for phenomenological applications. Notably, the large logarithmic terms are obtained analytically, and the explicit expressions established in this work provide a useful basis for further theoretical investigations.

We compute the unpolarized cross sections. The $\mathcal{O}(\alpha_s)$ correction is found to be large, while the $\mathcal{O}(\alpha_s^2)$ correction for $\chi_{c0}$ production amounts to $33\%$ of the leading-order (LO) cross section, significantly reducing the scale uncertainties. For $\chi_{c1}$, the $\mathcal{O}(\alpha_s)$ and $\mathcal{O}(\alpha_s^2)$ corrections correspond to $35\%$ and $-15\%$, respectively. For $\chi_{c2}$, the corresponding corrections are $24\%$ and $-38\%$. The large cancellation between the corrections for $\chi_{c2}$ brings the NNLO cross section close to the LO prediction. Our prediction for $\chi_{c0}$ is consistent with both the {\tt Belle} and {\tt BaBar} data within uncertainties.

We also predict the angular distribution parameters $\alpha^J_\theta$, which are independent of nonperturbative inputs. A sharp discrepancy between the theory and the {\tt Belle} measurement is observed for $\alpha^0_\theta$, calling for further experimental and theoretical investigations. Moreover, future measurements of the angular distribution parameters for $\chi_{c1}$ and $\chi_{c2}$ will provide important tests of the theoretical framework.
}

\begin{document}
\maketitle
\flushbottom

\section{Introduction}
\label{sec:intro}
\noindent 
Double charmonium production at B factories provides an important testing ground for understanding both perturbative and nonperturbative aspects of heavy quarkonium states. Following the first measurements by the {\tt Belle}~\cite{Belle:2002tfa} and {\tt BaBar}~\cite{BaBar:2005nic} collaborations, extensive theoretical studies have been devoted to this topic.

Significant discrepancies between LO nonrelativistic QCD (NRQCD) predictions~\cite{Braaten:2002fi,Hagiwara:2003cw,Liu:2002wq} and experimental data~\cite{Belle:2002tfa} for $e^+e^-\to J/\psi+\eta_c$ have attracted considerable theoretical attention. Subsequently, next-to-leading-order (NLO) QCD corrections~\cite{Zhang:2005cha,Gong:2007db}, relativistic corrections~\cite{Braaten:2002fi,He:2007te}, and mixed QCD and relativistic corrections~\cite{Dong:2012xx,Li:2013otv} have been computed. The NLO QCD and relativistic corrections were found
to be sizable, whereas the mixed $\mathcal{O}(\alpha_s v^2)$
corrections are comparatively small, indicating reasonably good
convergence of the combined perturbative and relativistic expansions~\cite{Li:2013otv}, where $v$ denotes the typical relative velocity between the charm quark and the anticharm quark inside charmonia, with $v^2\sim 0.3$. These corrections have significantly reduced the discrepancy between theory and experiment~\cite{Bodwin:2007ga}. 

The associated production of $J/\psi$ and $\chi_{cJ}$ ($J=0,1,2$) has also been measured. While the $J/\psi+\chi_{c0}$ channel has been clearly observed by both the {\tt Belle}~\cite{Belle:2004abn} and {\tt BaBar}~\cite{BaBar:2005nic} collaborations, the signals for $\chi_{c1}$ and $\chi_{c2}$ remain less conclusive, with only upper limits reported for their joint production rate~\cite{Belle:2004abn}. On the theory side, NLO QCD corrections have been computed in Refs.~\cite{Zhang:2008gp,Wang:2011qg,Dong:2011fb}. Large corrections for $J/\psi+\chi_{c0}$ production have been observed, improving the agreement with experimental data. Additional effects, such as renormalization scale dependence and QED interference, have been investigated in Refs.~\cite{Wang:2013vn,Jiang:2018wmv}. Furthermore, studies of angular distributions~\cite{Sun:2021tma} have revealed serious tensions between theoretical predictions and experimental measurements, calling for more precise calculations.

Different from the $J/\psi+\eta_c$ production, relativistic corrections to the $J/\psi+\chi_{cJ}$ processes are absent in the literature, due to the complexity of rigorous relativistic corrections for $P$-wave charmonia. For a rigorous perturbative matching calculation for $P$-wave quarkonium exclusive production, Ref.~\cite{Brambilla:2017kgw} computes the relativistic corrections to $e^+e^-\to \chi_{cJ}+\gamma$, where the calculation is already quite involved. Undoubtedly, the computation for $e^+e^-\to J/\psi+\chi_{cJ}$ is much more complicated. Although a precise estimation of the size of the relativistic corrections for this process is currently lacking, the uncalculated relativistic corrections may give rise to $\mathcal{O}(v^2)\sim 30\%$ corrections to the cross sections~\cite{Sang:2022kub}. In the present work, we concentrate on the QCD corrections.

With recent advances in multi-loop integral calculations, numerous NNLO QCD corrections to charmonium production at B factories have been computed~\cite{Chen:2017pyi,Sang:2020fql,Li:2025pbt,Feng:2019zmt,Sang:2022kub,Huang:2022dfw,Li:2025mng,Chen:2025qgy,Sang:2023liy,Huang:2023pmn}. In particular, the NNLO corrections to $e^+e^-\to J/\psi+\chi_{cJ}$ were already evaluated numerically for fixed center-of-mass (CM) energy and charm quark mass~\cite{Sang:2022kub}. Because these parameters are fixed, the resulting numerical results cannot be readily applied to predictions at different collision energies or with different choices of the charm quark mass. Moreover, pure numerical results do not allow one to study the asymptotic behavior as $\sqrt{s}\to \infty$, nor do they reveal information about the contributions of large logarithmic terms. Whether large logarithmic terms dominate the NNLO corrections remains an open question. An analytic expression, if obtainable, would not only greatly facilitate phenomenological applications—such as varying the CM energy or the charm quark mass—but also shed light on the resummation of large logarithms. Furthermore, it would offer the possibility to convert predictions from the on-shell mass scheme to other short-distance mass schemes, such as the $\overline{\rm MS}$ mass scheme. Unfortunately, deriving a full analytic expression for the cross sections of $e^+e^-\to J/\psi+\chi_{cJ}$ appears insurmountable due to the extremely complicated topology of the Feynman diagrams. As an alternative, our aim in this work is to obtain an asymptotic expansion in the variable $r={16m_c^2}/{s}$. This asymptotic expression includes various logarithmic terms and provides a good approximation to the exact results over the kinematic range $0\le r \le 1$.~\footnote{ We neglect relativistic corrections and therefore take the physical charmonium masses to be $2m_c$. In this treatment, $r=1$, i.e., $\sqrt{s}=4m_c$, corresponds to the threshold of double charmonium production. We emphasize, however, that the theoretical prediction based on NRQCD may become unreliable near the threshold $r=1$, since the interline gluons may become soft and spoil the perturbative expansion. Nevertheless, we retain the expansion around $r=1$ for completeness and for future theoretical interest.}

The remainder of this paper is organized as follows. In section~\ref{sec:theory}, we outline the theoretical framework for computing the cross sections and angular distribution parameters in terms of helicity amplitudes, and present the factorization of these amplitudes. In section~\ref{sec:tech}, we describe the computational methods used to evaluate the helicity amplitudes and to perform the asymptotic expansion of the master integrals. The convergence of the asymptotic expansion is examined in section~\ref{sec:convergent}. In section~\ref{sec:pheno}, we present phenomenological predictions for the cross sections and angular distribution parameters. Finally, a summary is given in section~\ref{sec:summary}. The helicity projectors are presented in appendix~\ref{app:hel:pro}. The asymptotic expansion coefficients up to $\mathcal{O}(r^0)$ are presented in appendix~\ref{app:asy:exp}. 

\section{Theoretical framework}\label{sec:theory}
\subsection{Helicity amplitudes and cross sections}
\noindent 
The exclusive process $e^+e^-\to J/\psi+\chi_{cJ}$ proceeds via the annihilation of the initial $e^+e^-$ pair into a timelike virtual photon, which subsequently decays into the charmonium pair. The differential cross section can therefore be expressed as~\cite{Sang:2022kub}
\begin{align}~\label{eq:diff:cross}
\frac{ d\sigma(J/\psi(\lambda_1)+\chi_{cJ}(\lambda_2))}{d\;\text{cos}\,\theta}&=\frac{2\pi\alpha}{s^{3/2}}\;\frac{d\Gamma(\gamma^*\to J/\psi(\lambda_1)+\chi_{cJ}(\lambda_2))}{d\;\cos\,\theta},
\end{align}
where $\lambda_1$ and $\lambda_2$ denote the helicities of the $J/\psi$ and $\chi_{cJ}$, respectively, and $\theta$ is the polar angle between the outgoing $J/\psi$ and the electron beam direction. Based on Refs.~\cite{Jacob:1959at,Haber:1994pe}, the differential decay width of the virtual photon is given by~\cite{Dong:2011fb}
\begin{align}
\frac{d\Gamma(\gamma^*\to J/\psi(\lambda_1)+\chi_{cJ}(\lambda_2))}{d\;\cos\,\theta}=\frac{\lvert \mathbf{P}\rvert}{16\pi s}\lvert\mathcal{A}^J_{\lambda_1,\lambda_2}\rvert^2\times\begin{cases} \frac{1+\text{cos}^2\theta}{2},&\lambda=\pm1 \\ 1-\text{cos}^2\theta,&\lambda=0\quad,\label{eq:1}
\end{cases}
\end{align}
where $\sqrt{s}$ is the CM energy, and $\lambda=\lambda_1-\lambda_2$. Note that angular momentum conservation restricts $\lvert \lambda \rvert \leq 1$.
$\lvert \mathbf{P}\rvert$ denotes the magnitude of the three-momentum of the $J/\psi$  in the CM frame
\begin{equation}~\label{eq:mom}
    \lvert \mathbf{P}\rvert=\sqrt{\frac{\lambda(s,M_{J/\psi}^2,M_{\chi_{cJ}}^2)}{4s}},
\end{equation}
with $\lambda(x,y,z)=x^2+y^2+z^2-2xy-2xz-2yz$ being the K\"allen function.

By parity invariance, the helicity amplitudes satisfy
\begin{equation}\label{eq:parity}
    \mathcal{A}_{\lambda_1,\lambda_2}^J=(-1)^J\mathcal{A}_{-\lambda_1,-\lambda_2}^J,
\end{equation}
which implies that there are 10 independent amplitudes: 2 for $\chi_{c0}$ production, 3 for $\chi_{c1}$ and 5 for $\chi_{c2}$. In particular, Eq.~(\ref{eq:parity}) forces $\mathcal{A}^1_{0,0}=0$.

Summing over all helicities of the final particles in Eq.~\eqref{eq:diff:cross}, the unpolarized differential cross section becomes
\begin{equation}
    \frac{d\sigma(J/\psi+\chi_{cJ})}{d\;\text{cos}\,\theta}=\frac{\alpha}{8s^2}\left(\frac{\lvert\mathbf{P}\rvert}{\sqrt{s}}\right)A^J(1+\alpha_\theta^J\,\text{cos}^2\,\theta),\qquad J=0,1,2\label{eq:unpol:cross}
\end{equation}
where $\alpha_\theta^J$ is a dimensionless parameter governing the angular distribution, with $\lvert\alpha_\theta^J\rvert\le 1$.
From Eq.~\eqref{eq:1} and Eq.~\eqref{eq:unpol:cross}, one obtains
\begin{equation}\label{eq:eqa}
A^0=\lvert\mathcal{A}_{1,0}^0\rvert^2+\lvert\mathcal{A}_{0,0}^0\rvert^2,\qquad \alpha^0_\theta=-\frac{\lvert\mathcal{A}_{0,0}^0\rvert^2-\lvert\mathcal{A}_{1,0}^0\rvert^2}{\lvert\mathcal{A}_{0,0}^0\rvert^2+\lvert\mathcal{A}_{1,0}^0\rvert^2}
\end{equation}
for $J/\psi+\chi_{c0}$,
\begin{equation}
A^1=\lvert\mathcal{A}_{1,0}^1\rvert^2+\lvert\mathcal{A}_{0,1}^1\rvert^2+2\lvert\mathcal{A}_{1,1}^1\rvert^2,\qquad \alpha^1_\theta=\frac{\lvert\mathcal{A}_{1,0}^1\rvert^2+\lvert\mathcal{A}_{0,1}^1\rvert^2-2\lvert\mathcal{A}_{1,1}^1\rvert^2}{\lvert\mathcal{A}_{1,0}^1\rvert^2+\lvert\mathcal{A}_{0,1}^1\rvert^2+2\lvert\mathcal{A}_{1,1}^1\rvert^2}
\end{equation}
for $J/\psi+\chi_{c1}$,
\begin{subequations}
\begin{eqnarray}
A^2&=&\lvert\mathcal{A}_{1,0}^2\rvert^2+\lvert\mathcal{A}_{0,1}^2\rvert^2+2\lvert\mathcal{A}_{1,1}^2\rvert^2+\lvert\mathcal{A}_{1,2}^2\rvert^2+\lvert\mathcal{A}_{0,0}^2\rvert^2,\\
\alpha^2_\theta&=&-\frac{\lvert\mathcal{A}_{0,0}^2\rvert^2-\lvert\mathcal{A}_{0,1}^2\rvert^2+2\lvert\mathcal{A}_{1,1}^2\rvert^2-\lvert\mathcal{A}_{1,0}^2\rvert^2-\lvert\mathcal{A}_{1,2}^2\rvert^2}{\lvert\mathcal{A}_{0,0}^2\rvert^2+\lvert\mathcal{A}_{0,1}^2\rvert^2+2\lvert\mathcal{A}_{1,1}^2\rvert^2+\lvert\mathcal{A}_{1,0}^2\rvert^2+\lvert\mathcal{A}_{1,2}^2\rvert^2},
\end{eqnarray}
\end{subequations}
for $J/\psi+\chi_{c2}$.

Finally, integrating over $\theta$ yields the total unpolarized cross sections:
\begin{subequations}
\begin{eqnarray}
    \sigma(J/\psi+\chi_{c0})&=&\frac{\alpha}{6s^2}\frac{\lvert\mathbf{P}\rvert}{\sqrt{s}}\bigg(2\lvert\mathcal{A}^0_{1,0}\rvert^2+\lvert\mathcal{A}^0_{0,0}\rvert^2\bigg),\\
    \sigma(J/\psi+\chi_{c1})&=&\frac{\alpha}{6s^2}\frac{\lvert\mathbf{P}\rvert}{\sqrt{s}}\bigg(2\lvert\mathcal{A}^1_{1,0}\rvert^2+2\lvert\mathcal{A}^1_{0,1}\rvert^2+2\lvert\mathcal{A}^1_{1,1}\rvert^2\bigg), \\
    \sigma(J/\psi+\chi_{c2})&=&\frac{\alpha}{6s^2}\frac{\lvert\mathbf{P}\rvert}{\sqrt{s}}\bigg(2\lvert\mathcal{A}^2_{1,0}\rvert^2+2\lvert\mathcal{A}^2_{0,1}\rvert^2+2\lvert\mathcal{A}^2_{1,1}\rvert^2+2\lvert\mathcal{A}^2_{1,2}\rvert^2+\lvert\mathcal{A}^2_{0,0}\rvert^2\bigg).
\end{eqnarray}
\end{subequations}

\subsection{NRQCD factorization for the helicity amplitudes}\label{sec:2.2}

We apply the NRQCD factorization~\cite{Bodwin:1994jh} to express the helicity amplitudes as
\begin{equation}
\mathcal{A}^J_{\lambda_1,\lambda_2}=c^J_{\lambda_1,\lambda_2}\sqrt{2M_{J/\psi}}\sqrt{2M_{\chi_{cJ}}}\frac{\langle\mathcal{O}_{^3S_1}\left(\mu_\Lambda\right)\rangle\,\langle\mathcal{O}_{^3P_J}\left(\mu_\Lambda\right)\rangle}{m_c^4}\bigg(1+\mathcal{O}(v^2)\bigg),\label{eq:nrqcd}
\end{equation}
where $c^J_{\lambda_1,\lambda_2}$ denotes the short-distance coefficient (SDC). The prefactors $\sqrt{2M_{J/\psi}}$ and $\sqrt{2M_{\chi_{cJ}}}$ appear because we employ the relativistic normalization for quarkonium states in the helicity amplitudes, while the long-distance matrix elements (LDMEs) adopt the nonrelativistic normalization. The LDMEs are defined as
\begin{subequations}
\begin{eqnarray}
    \langle\mathcal{O}_{^3S_1}(\mu_\Lambda)\rangle\;&=&\;\langle J/\psi\lvert\psi^\dagger\boldsymbol{\sigma}\cdot\epsilon_{J/\psi}\,\chi(\mu_{\Lambda})\rvert0\rangle,\\
    \langle\mathcal{O}_{^3P_0}(\mu_\Lambda)\rangle\;&=&\;\langle \chi_{c0}\lvert\ \psi^\dagger\frac{1}{\sqrt{3}}\left(-\frac{i}{2}\boldsymbol{\overleftrightarrow{D}\cdot\sigma}\right)\,\chi(\mu_{\Lambda})\rvert0\rangle,\\
    \langle\mathcal{O}_{^3P_1}(\mu_\Lambda)\rangle\;&=&\;\langle \chi_{c1}\lvert\ \psi^\dagger\frac{1}{\sqrt{2}}\left(-\frac{i}{2}\boldsymbol{\overleftrightarrow{D}\times\sigma}\right)\cdot\boldsymbol{\epsilon}_{\chi_{c1}}\,\chi(\mu_{\Lambda})\rvert0\rangle,\\
    \langle\mathcal{O}_{^3P_2}(\mu_\Lambda)\rangle\;&=&\;\langle \chi_{c2}\lvert\ \psi^\dagger\left(-\frac{i}{2}\overleftrightarrow{D}{}^{(i}\sigma^{j)}\epsilon^{ij}_{\chi_{c2}}\right)\,\chi(\mu_{\Lambda})\rvert0\rangle,
\end{eqnarray}
\end{subequations}
where $\psi^{\dagger}$ and $\chi$ are the Pauli spinor fields that create a charm quark and an anticharm quark, respectively,
$\epsilon_{J/\psi}$ and $\epsilon_{\chi_{c1}}$ denote the polarization vectors of the $J/\psi$ and $\chi_{c1}$, while $\epsilon_{\chi_{c2}}$ is the polarization tensor of the $\chi_{c2}$. The scale $\mu_\Lambda$ is the factorization scale. The $\mu_\Lambda$-dependence in the LDMEs is canceled by the corresponding dependence in the SDCs, rendering the helicity amplitudes $\mu_\Lambda$-independent.

Because SDCs are insensitive to nonperturbative hadronization effects, they can be determined through the standard perturbative matching procedure. Specifically, we replace the physical $J/\psi$ and $\chi_{cJ}$ in the final state by fictitious onium states, each consisting of a free $c\bar{c}$ pair carrying the same quantum numbers as the corresponding physical charmonium. After this replacement, Eq.~\eqref{eq:nrqcd} becomes
\begin{equation}
\mathcal{A}^{J}_{\lambda_1,\lambda_2}(c\bar{c}(^3S_1),c\bar{c}(^3P_J))=c^{J}_{\lambda_1,\lambda_2}\frac{\langle\mathcal{O}_{c\bar{c}(^3S_1)}\rangle\,\langle\mathcal{O}_{c\bar{c}(^3P_J)}\rangle}{m_c^4},\label{eq:nrqcd:per}
\end{equation}
where the prefactors $\sqrt{2M_{J/\psi}}$ and $\sqrt{2M_{\chi_{cJ}}}$ have been omitted because we adopt the relativistic normalization for the charm-quark states on both sides of Eq.~\eqref{eq:nrqcd:per}.
With this replacement, both sides of the equation can be computed perturbatively, order by order in $\alpha_s$.

Explicitly, 
the perturbative LDMEs are given by
\begin{subequations}
\begin{align}
&\langle\mathcal{O}_{c\bar{c}(^3S_1)}\rangle\bigg\rvert_{\overline{\rm MS}}=2m_c\sqrt{2N_c}
\bigg(1+\frac{\alpha_s^2}{\pi^2}\frac{1}{2\epsilon_{\rm IR}}\gamma_{J/\psi}\bigg)+\mathcal{O}(\alpha_s^3),
\\
&\langle\mathcal{O}_{c\bar{c}(^3P_J)}\rangle\bigg\rvert_{\overline{\rm MS}}=2m_c\sqrt{2N_c}\bigg(1+\frac{\alpha_s^2}{\pi^2}\frac{1}{2\epsilon_{\rm IR}}\gamma_{\chi_{cJ}}\bigg)+\mathcal{O}(\alpha_s^3), 
\end{align}
\end{subequations}
where $N_c=3$, and $\gamma_{J/\psi}$ and $\gamma_{\chi_{cJ}}$ are the anomalous dimensions associated with the NRQCD bilinear currents for the $^3S_1$ and $^3P_J$ states, respectively, which are given by \cite{Czarnecki:1997vz,Beneke:1997jm,Hoang:2006ty,Sang:2015uxg,Sang:2020fql}:
\begin{subequations}
\begin{align}
    &\gamma_{J/\psi}\;=\;-\pi^2\left(\frac{C_AC_F}{4}+\frac{C_F^2}{6}\right),\\&
    \gamma_{\chi_{c0}}\;=\;-\pi^2\left(\frac{C_AC_F}{12}+\frac{C_F^2}{3}\right),\\&
     \gamma_{\chi_{c1}}\;=\;-\pi^2\left(\frac{C_AC_F}{12}+\frac{5C_F^2}{24}\right),\\&
      \gamma_{\chi_{c2}}\;=\;-\pi^2\left(\frac{C_AC_F}{12}+\frac{13C_F^2}{120}\right).
\end{align}
\end{subequations}

To compute the left-hand side of Eq.~\eqref{eq:nrqcd:per}, we first generate the amplitude for $\gamma^*\to c\bar{c}(P_1)+c\bar{c}(P_2)$, where $P_1$ and $P_2$ denote the momenta of the charm pairs, respectively. The color-singlet/spin-triplet states can be extracted by employing the relativistically normalized projectors~\cite{Petrelli:1997ge,Bodwin:2002cfe}
\begin{subequations}
    \begin{align}
\Pi_{J/\psi}&=-\frac{1}{8\sqrt{2}m_c^2}(\slashed{\bar{p}}_1-m_c)\slashed{\epsilon}_{J/\psi}^*(\slashed{P}_1+2m_c)(\slashed{p}_1+m_c)\otimes \frac{\mathbf{1}_c}{\sqrt{N_c}},\\
\Pi^\alpha_{\chi_{cJ}}&=-\frac{1}{8\sqrt{2}m_c^2}(\slashed{\bar{p}}_2-m_c)\gamma^\alpha(\slashed{P}_2+2m_c)(\slashed{p}_2+m_c)\otimes \frac{\mathbf{1}_c}{\sqrt{N_c}},
    \end{align}
\end{subequations}
where   
\begin{subequations}
\begin{align}
 &p_1=\bar{p}_1=\frac{P_1}{2},\\
 &   p_2=\frac{P_2}{2}+q_2,\;\bar{p}_2=\frac{P_2}{2}-q_2,
\end{align}
\end{subequations}
with $q_2$ denoting the relative momentum between the charm quark and the anti-charm quark of the $c\bar{c}(P_2)$ pair. 

The color-singlet spin-triplet quark-level amplitude can be obtained by replacing the charm spinors $v(\bar{p}_1)\bar{u}(p_1)$ with $\Pi_{J/\psi}$ and $v(\bar{p}_2)\bar{u}(p_2)$ with $\Pi_{\chi_{cJ}}^\alpha$ from the amplitude of $\gamma^*\to c\bar{c}(P_1)+c\bar{c}(P_2)$.
Furthermore, the amplitude for $\gamma^*\to c\bar{c}(^3S_1)+c\bar{c}(^3P_J)$
is then obtained by differentiating the color-singlet spin-triplet amplitude with respect to the relative momentum $q_2$:
 \begin{align}
 \label{eq:per:pol}
 \mathcal{A}^{J}(c\bar{c}(^3S_1),c\bar{c}(^3P_J))=\mathcal{Q}_{\alpha\beta}^{J}\,\frac{\partial}{\partial q_2^\beta}\text{Tr}[\Pi^\alpha_{\chi_{cJ}} \mathcal{A}\,\Pi_{J/\psi}]\bigg\rvert_{q_2=0},
 \end{align}
where $\mathcal{A}$ denotes the amplitude for $\gamma^*\to c\bar{c}+c\bar{c}$ with the external quark spinors truncated, and the projection operators $\mathcal{Q}_{\alpha\beta}^{J}$ are constructed from the Clebsch–Gordan coefficients:
\begin{subequations}
    \begin{align}
        \mathcal{Q}_{\alpha\beta}^{0}\;&=\;\frac{-1}{\sqrt{3}}(-g_{\alpha\beta}+\frac{P_{2\alpha} P_{2\beta}}{4m_c^2}),\\
        \mathcal{Q}_{\alpha\beta}^{1}\;&=\;\frac{i}{2\sqrt{2}m_c}\epsilon_{\alpha\beta\rho\sigma}\,P_2^\rho\epsilon^{*\sigma}_{\chi_{c1}},\\
\mathcal{Q}_{\alpha\beta}^{2}\;&=\;\epsilon^{*}_{\chi_{c2},\alpha\beta}.
    \end{align}
\end{subequations}

To further obtain the helicity amplitudes $\mathcal{A}^{J}_{\lambda_1,\lambda_2}(c\bar{c}(^3S_1),c\bar{c}(^3P_J))$ from Eq.~\eqref{eq:per:pol}, it is very useful to construct various helicity projectors $\mathcal{P}_{\lambda_1,\lambda_2}^{(\chi_{cJ})}$. The construction, explicit expressions, and application of these projectors to extract the helicity amplitudes are presented in Appendix~\ref{app:hel:pro}.

With all the necessary ingredients at hand, we are able to evaluate both sides of Eq.~\eqref{eq:nrqcd:per} order by order in perturbation theory, thereby determining the SDCs in a systematic manner. For further details, we refer the reader to Refs.~\cite{Sang:2022kub,Zhang:2021ted}.

\subsection{SDCs in $\alpha_s$ expansion}
\noindent The SDCs can be expressed up to $\mathcal{O}(\alpha_s^2)$ as
\begin{align}
c^J_{\lambda_1,\lambda_2}&(r,\mu_\Lambda)=\frac{16\pi e\alpha_s}{27\sqrt{3}}\bigg(\frac{r}{4}\bigg)^{(1+\lvert\lambda_1+\lambda_2\rvert)/2}\bigg\{c^{J\,{(0)}}_{\lambda_1,\lambda_2}+\frac{\alpha_s(\mu_R)}{\pi}\left(c^{J\,{(0)}}_{\lambda_1,\lambda_2}\frac{\beta_0}{4}\ln\frac{4\mu_R^2}{s}+c^{J\,{(1)}}_{\lambda_1,\lambda_2}\right)\notag \\& \notag+\frac{\alpha_s^2(\mu_R)}{\pi^2}\bigg[c^{J\,{(0)}}_{\lambda_1,\lambda_2}\bigg(\frac{\beta_0^2}{16}\ln^2\frac{4\mu_R^2}{s}+\frac{\beta_1}{16}\ln\frac{4\mu_R^2}{s}\bigg)+c^{J\,{(1)}}_{\lambda_1,\lambda_2}\frac{\beta_0}{2}\ln\frac{4\mu_R^2}{s}
\\& +c^{J\,{(0)}}_{\lambda_1,\lambda_2}\left(\gamma_{J/\psi}+\gamma_{\chi_{cJ}}\right)\ln\frac{\mu_\Lambda^2}{m_c^2}+c^{J\,{(2)}}_{\lambda_1,\lambda_2}\bigg]\bigg\},
\end{align}
where $r = {16m_c^2}/{s}$. The first two coefficients of the QCD $\beta$ function are $\beta_0={11C_A}/{3}-{4T_Fn_f}/{3}$, $\beta_1={34C_A^2}/{3}-{20C_AT_Fn_f}/{3}-4C_FT_Fn_f$, where $n_f=n_l+n_h$ denotes the number of active quark flavors. Here $n_l=3$ is the number of light quarks, and $n_h=1$ is the number of heavy quarks. Since we do not include contributions from bottom quark loops, $\alpha_s$ should be evaluated with $n_f=4$. 

$\mu_R$ and $\mu_\Lambda$ denote the renormalization scale and the NRQCD factorization scale, respectively. The presence of the $\ln \mu_R$ terms ensures the renormalization-group invariance of $c^J_{\lambda_1,\lambda_2}$ at two-loop accuracy, while the $\ln \mu_\Lambda$ term arises from the factorization requirement.
The $\mu_\Lambda$-dependence in $c^J_{\lambda_1,\lambda_2}$ is canceled by the corresponding dependence in the LDMEs.

The tree-level SDCs $c^{J\,(0)}_{\lambda_1,\lambda_2}$ have been known for a long time~\cite{Braaten:2002fi,Dong:2011fb}:
\begin{subequations}
    \begin{equation}
        c^{0\,(0)}_{1,0}\;=\;9-\frac{7}{2}r,\; c^{0\,(0)}_{0,0}\;=\;\frac{1}{4}(4+10r-3r^2),
    \end{equation}
    \begin{equation}
        c^{1\,(0)}_{1,0}\;=\;-\frac{\sqrt{6}}{4}r,\;
         c^{1\,(0)}_{0,1}\;=\;-\frac{\sqrt{6}}{4}(8-7r),\;
          c^{1\,(0)}_{1,1}\;=\; -\frac{\sqrt{6}}{2}(4-3r),
    \end{equation}
    \begin{equation}
         c^{2\,(0)}_{1,2}\;=\;-2\sqrt{3},\; c^{2\,(0)}_{1,0}\;=\;-\frac{\sqrt{2}}{4}(12-11r),\;
         c^{2\,(0)}_{1,1}\;=\;-\frac{\sqrt{6}}{2}(4-3r),
    \end{equation}
    \begin{equation}
         c^{2\,(0)}_{0,1}\;=\;-\frac{\sqrt{6}}{4}(4-5r),\;
          c^{2\,(0)}_{0,0}\;=\;-\frac{\sqrt{2}}{4}(4-2r-3r^2).
    \end{equation}
\end{subequations}

The $\mathcal{O}(\alpha_s)$ coefficients $c^{J(1)}_{\lambda_1,\lambda_2}$ have been computed in Ref.~\cite{Dong:2011fb}.~\footnote{The $\mathcal{O}(\alpha_s)$ corrections to the unpolarized cross sections for $e^+e^- \to J/\psi + \chi_{cJ}$ can also be found in Refs.~\cite{Zhang:2008gp,Wang:2011qg}.} 
The $\mathcal{O}(\alpha_s^2)$ coefficients $c^{J(2)}_{\lambda_1,\lambda_2}$ were evaluated numerically in Ref.~\cite{Sang:2022kub}. The primary goal of this work is to derive semi-analytical expressions for the $\mathcal{O}(\alpha_s^2)$ corrections to $c^{J(2)}_{\lambda_1,\lambda_2}$.

\section{Computational technologies}\label{sec:tech}
\begin{figure} [t]
    \centering
    \includegraphics[width=1\linewidth]{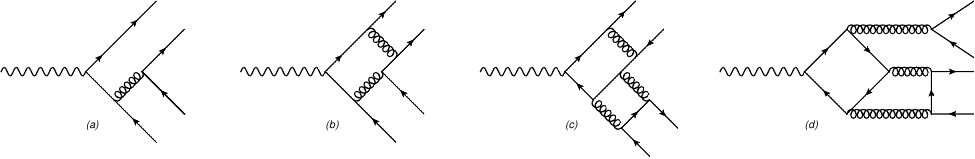}
    \caption{Representative Feynman diagrams for $\gamma^*\to c\bar{c}+c\bar{c}$ at LO (a), NLO (b), and NNLO (c,d), where (d) denotes the ``light-by-light" (lbl) contribution. Drawn with \texttt{JaxoDraw}~\cite{Binosi:2008ig}.}
    \label{fig:feynmanpic}
\end{figure}

\noindent 
We employ {\tt FeynArts}~\cite{Hahn:2000kx} to generate the Feynman diagrams and the corresponding amplitudes for the process $\gamma^*\to c\bar{c}+c\bar{c}$ through $\mathcal{O}(\alpha_s^2)$ QCD corrections. Representative diagrams are shown in Fig.~\ref{fig:feynmanpic}. Since the sum of the electric charges of the light quarks vanishes, i.e., $e_u+e_d+e_s=0$, the ``light-by-light” contributions  from light-quark loops can be safely neglected.

Following the procedure described in Sec.~\ref{sec:2.2}, we apply a set of projectors—including spin/color, orbital-spin coupling, and helicity projectors—to extract the desired helicity amplitudes. The Dirac algebra and Lorentz contractions are performed using {\tt FeynCalc} and {\tt FormLink}~\cite{Mertig:1990an,Feng:2012tk}.

For the loop calculations, we use {\tt CalcLoop}~\cite{calcloop} to classify the Feynman integrals into distinct families. These integrals are then reduced to master integrals (MIs) via integration-by-parts (IBP) identities, employing the auxiliary mass flow package {\tt AMFlow}~\cite{Liu:2017jxz,Liu:2022chg,Liu:2022mfb} in conjunction with {\tt Kira}~\cite{Klappert:2020nbg} and {\tt Blade}~\cite{Guan:2024byi}. We also use {\tt FIRE}~\cite{Smirnov:2014hma} to identify symmetries among MIs originating from different families. Ultimately, we obtain 7 integral families and 13 master integrals at one loop, and 58 families with 463 MIs at two loops.

All calculations are carried out in $d = 4 - 2\epsilon$ dimensions.
The asymptotic expansions of the MIs around $r=r_0$ are obtained by solving their corresponding differential equations (DEs) with respect to $r$.  
For a complete set of MIs $\vec{J}$ within a given family, the DEs take the form
\begin{equation}\label{eq-de}
    \frac{\partial}{\partial r}\vec{J}\;=\;M(r,\epsilon)\vec{J},
\end{equation}
where $M(r,\epsilon)$ is Fuchsian at $r=r_0$, meaning that it has at most simple poles at this point. This structure guarantees that the solution admits an expansion of the form
\begin{equation}\label{eq-asy}
    \vec{J}(r,\epsilon) = \sum_{\alpha} (r-r_0)^\alpha \sum_{k=0}^{k_\alpha}\sum_{n=0}^{\infty} \vec{a}_{\alpha,k,n}(\epsilon)\,(r-r_0)^{n}\ln^k (r-r_0),
\end{equation}
where $\alpha$ is a finite set of  $\epsilon$-dependent characteristic exponents determined by the DEs, and $k_\alpha$ is an integer indicating the maximal power of $\ln(r-r_0)$ associated with the exponent $\alpha$. Substituting Eq.~\eqref{eq-asy} into Eq.~\eqref{eq-de}, we determine the exponents $\alpha$ by matching the lowest-order terms and compute the coefficients $\vec{a}$ from recurrence relations and boundary conditions. The boundary conditions are obtained from the high-precision numerical values of the MIs evaluated at a regular point close to $r_0$ using the package {\tt AMFlow}.

In practice, we adopt a numerical strategy. Specifically, we first assign $\epsilon$ to a set of small rational numbers and obtain the corresponding asymptotic expansion for each value. The final  $\epsilon$-dependent expansion is then reconstructed via a fitting procedure. 
After performing the expansion around $\epsilon=0$, the master integrals, and hence the helicity amplitudes, can be expressed as series expansions in both $\epsilon$ and $r-r_0$.
These operations are performed using the {\tt AMFlow} package. This technique has been employed in our previous studies~\cite{Li:2025mng,Li:2025pbt} and also reported in Refs.~\cite{Chen:2025qgy,Chen:2026maw}.

For renormalization, we adopt the on-shell scheme for the charm-quark mass and field strength, and the $\overline{\rm{MS}}$ scheme for the QCD coupling constant to remove ultraviolet divergences. A residual infrared divergence remains; its coefficient is exactly half the sum of the anomalous dimensions of the $J/\psi$ and $\chi_{cJ}$ \cite{Czarnecki:1997vz,Beneke:1997jm,Hoang:2006ty,Sang:2015uxg,Sang:2020fql}. Consequently, it can be completely absorbed into the corresponding NRQCD LDMEs, rendering the SDCs finite.

Finally, the SDCs can be expressed as a series of the form 
\begin{equation}\label{eq:ff}
    c^{J\,(n)}_{\lambda_1,\lambda_2}=\sum_{i,j}f^{J\,(n)}_{\lambda_1,\lambda_2}(i,j)(r-r_0)^i\ln^j(r-r_0),
\end{equation}
For the expansion around $r_0=0$, the high-precision numerical values of the coefficients $f^{J\,(n)}_{\lambda_1,\lambda_2}(i,j)$ allow us to reconstruct their analytic expressions for lower powers of $r$ using the {\tt PSLQ} algorithm. Some expressions are presented in appendices~\ref{app1} and \ref{app2}.

For the convenience of readers and for phenomenological applications, we provide the complete expansion coefficients at both $r_0=0$ and $r_0=1$, up to $\mathcal{O}(r^{20})$, in the electronic supplementary material files. Notably, the full color-factor dependence in the expansion coefficients is explicitly retained. In addition, we numerically compute the SDCs at 619 values of $r$ covering the interval $0< r < 1$ using {\tt AMFlow} with high precision. To distinguish these from the asymptotic expansions, we refer to these numerical values as the exact numerical results in this work. These results are also included in the attached electronic supplementary material files.

\section{Confronting the NNLO asymptotic expressions with the exact results}\label{sec:convergent}
It is instructive to examine the convergence of the asymptotic expansion by comparing the asymptotic expressions with the exact numerical results for the SDCs. We adopt the asymptotic expansion around $r=0$ in section~\ref{subsec:r=0}, and the combined expansion around both $r=0$ and $r=1$ in section~\ref{subsec:r=0:r=1}.

\subsection{Convergence behavior of the asymptotic expansion at $r=0$~\label{subsec:r=0}}
\noindent 
Prior to the comparison, it is convenient to introduce a shorthand notation:
\begin{equation}\label{eq:4.1}
    g_{i}=\sum^i_{a=-1}\sum^4_{b=0} f^{J\,(2)}_{\lambda_1,\lambda_2}(a,b)\,r^a\,\ln^b r,
\end{equation}
where, for brevity, we have suppressed the indices $J$ (representing $J/\psi+\chi_{cJ}$ final state) and the helicities $\lambda_1$ and $\lambda_2$ in $g$. 
Consequently, $g_{i}$ implicitly refers to a specific $J/\psi+\chi_{cJ}$ with helicities $\lambda_1$ and $\lambda_2$ when comparing the asymptotic expression for $c^{J\,(2)}_{\lambda_1,\lambda_2}$ with its exact values. 
Similarly, the exact numerical results are denoted by $g_{\rm exact}$. It is worth noting that the maximal power of logarithms in Eq.~\eqref{eq:4.1} is dictated by the infrared structure of the amplitudes, which is governed by soft and collinear divergences in the limit $\sqrt{s}\gg m_c$.

\begin{figure} [htbp]
    \centering
    \begin{minipage}{0.48\textwidth}
        \centering
        \begin{subfigure}{\linewidth}
            \includegraphics[width=\linewidth]{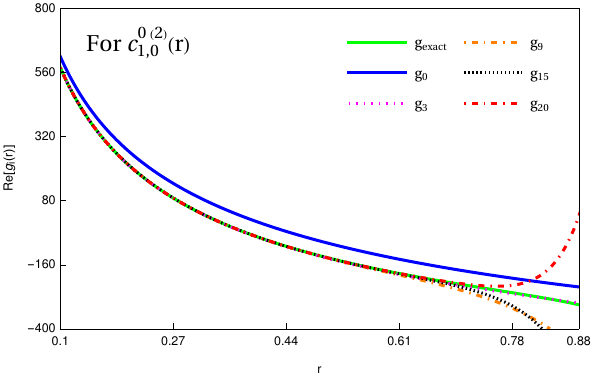}
        \end{subfigure}
        \begin{subfigure}{\linewidth}
            \includegraphics[width=\linewidth]{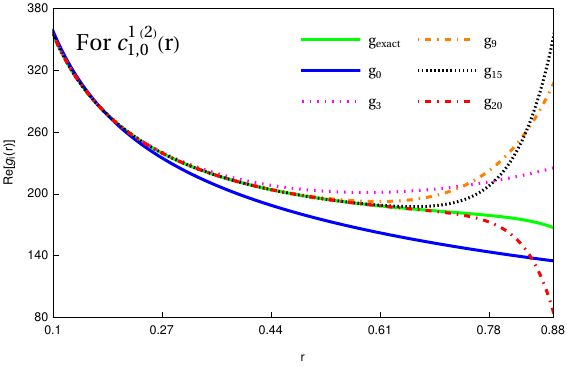}
        \end{subfigure}
        \begin{subfigure}{\linewidth}
            \includegraphics[width=\linewidth]{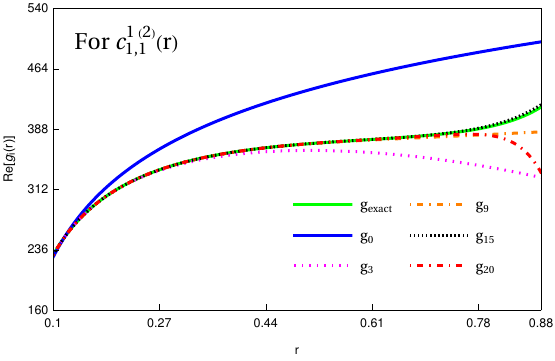}
        \end{subfigure}
        \begin{subfigure}{\linewidth}
            \includegraphics[width=\linewidth]{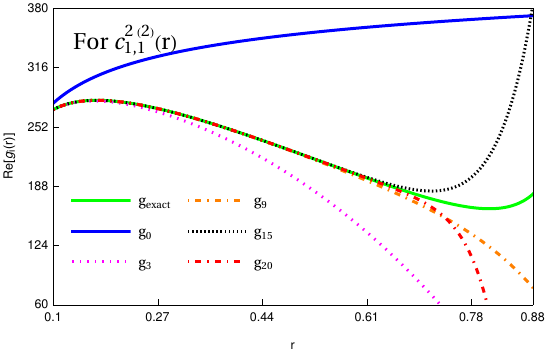}
            \begin{subfigure}{\linewidth}
            \includegraphics[width=\linewidth]{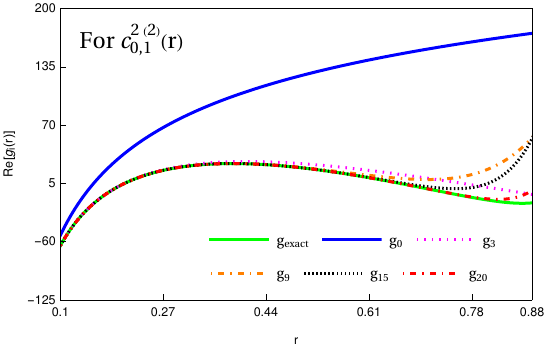}
        \end{subfigure}
        \end{subfigure}
    \end{minipage}
    \hspace{0.01\textwidth}
    \begin{minipage}{0.48\textwidth}
        \centering
        \begin{subfigure}{\linewidth}
            \includegraphics[width=\linewidth]{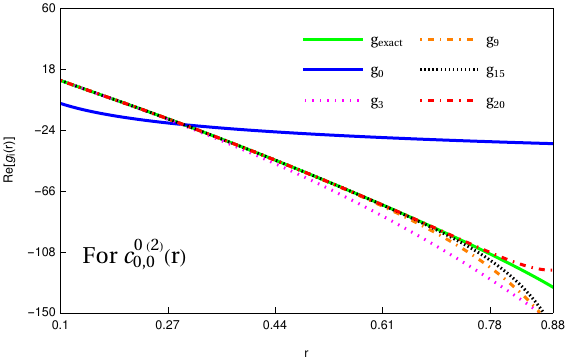}
        \end{subfigure}
        \begin{subfigure}{\linewidth}
            \includegraphics[width=\linewidth]{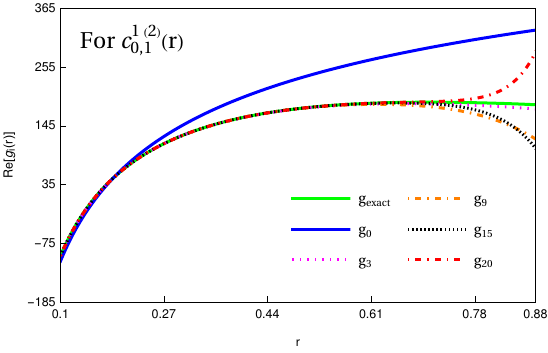}
        \end{subfigure}
        \begin{subfigure}{\linewidth}
            \includegraphics[width=\linewidth]{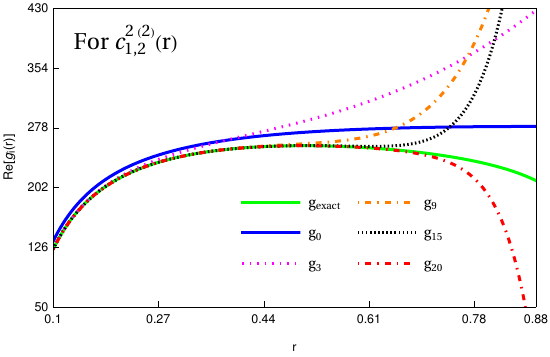}
        \end{subfigure}
        \begin{subfigure}{\linewidth}
            \includegraphics[width=\linewidth]{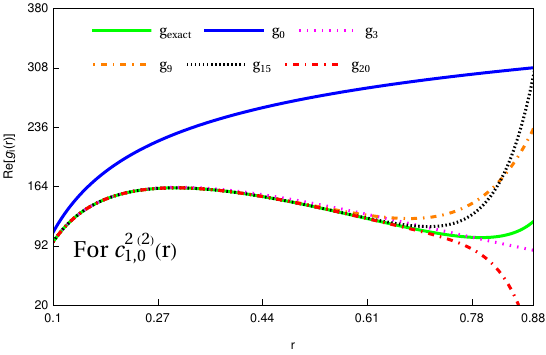}
            \begin{subfigure}{\linewidth}
            \includegraphics[width=\linewidth]{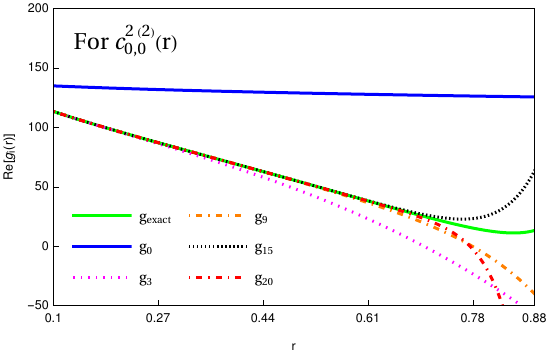}
        \end{subfigure}
        \end{subfigure}
    \end{minipage}
    \caption{Comparison between the real parts of the asymptotic expansions around $r=0$ and the exact numerical results.}
    \label{fig:re}
\end{figure}

\begin{figure} [htbp]
    \centering
    \begin{minipage}{0.48\textwidth}
        \centering
        \begin{subfigure}{\linewidth}
            \includegraphics[width=\linewidth]{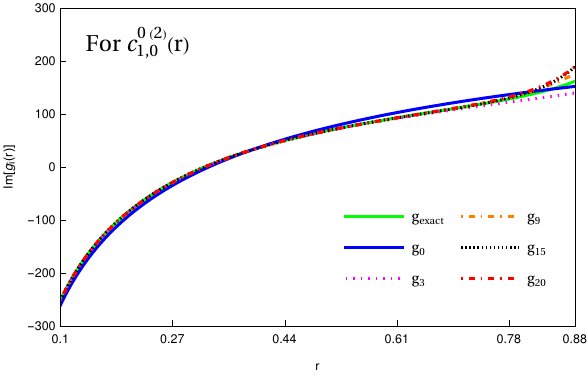}
        \end{subfigure}
        \begin{subfigure}{\linewidth}
            \includegraphics[width=\linewidth]{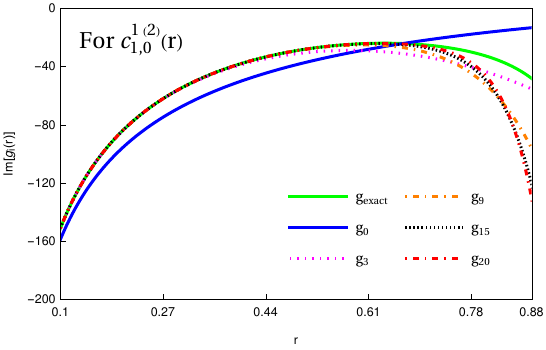}
        \end{subfigure}
        \begin{subfigure}{\linewidth}
            \includegraphics[width=\linewidth]{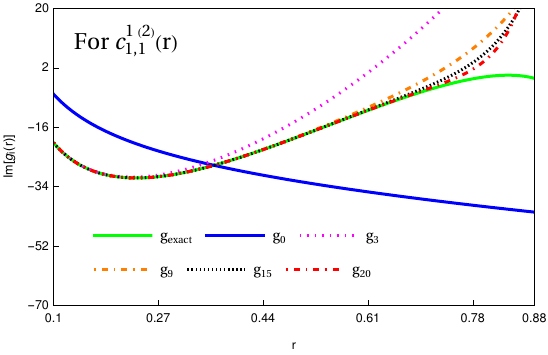}
        \end{subfigure}
        \begin{subfigure}{\linewidth}
            \includegraphics[width=\linewidth]{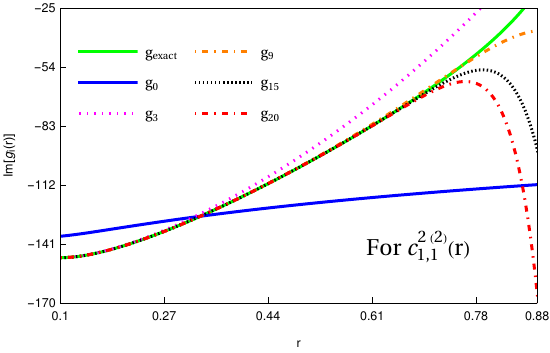}
            \begin{subfigure}{\linewidth}
            \includegraphics[width=\linewidth]{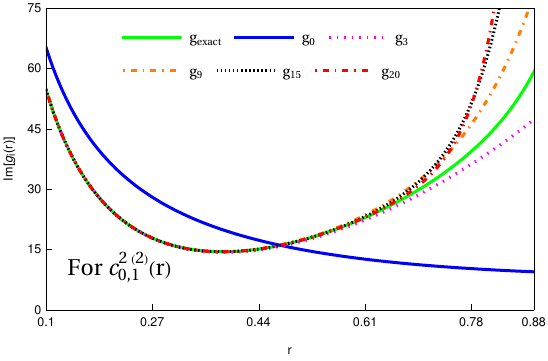}
        \end{subfigure}
        \end{subfigure}
    \end{minipage}
    \hspace{0.01\textwidth}
    \begin{minipage}{0.48\textwidth}
        \centering
        \begin{subfigure}{\linewidth}
            \includegraphics[width=\linewidth]{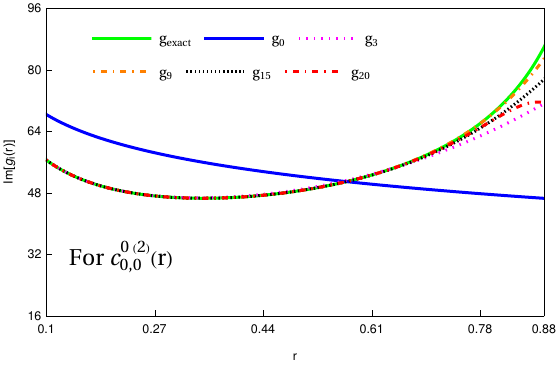}
        \end{subfigure}
        \begin{subfigure}{\linewidth}
            \includegraphics[width=\linewidth]{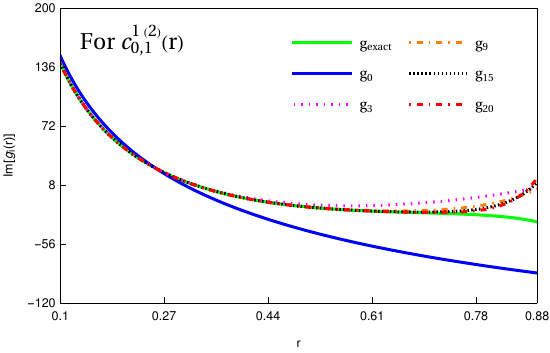}
        \end{subfigure}
        \begin{subfigure}{\linewidth}
            \includegraphics[width=\linewidth]{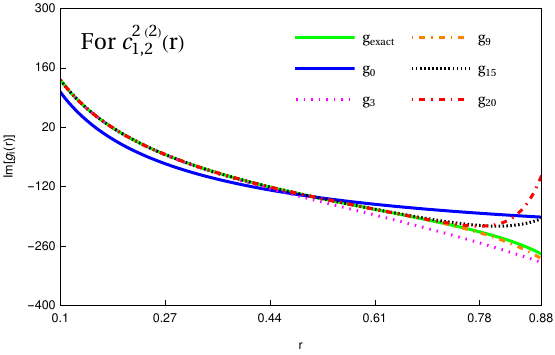}
        \end{subfigure}
        \begin{subfigure}{\linewidth}
            \includegraphics[width=\linewidth]{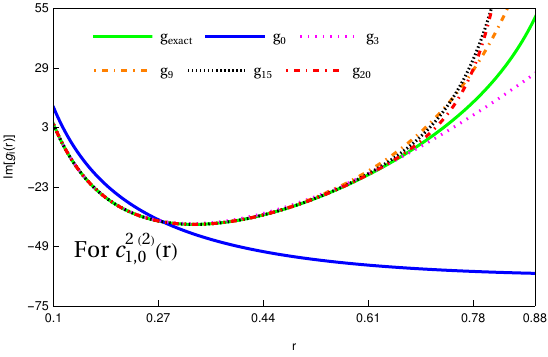}
            \begin{subfigure}{\linewidth}
            \includegraphics[width=\linewidth]{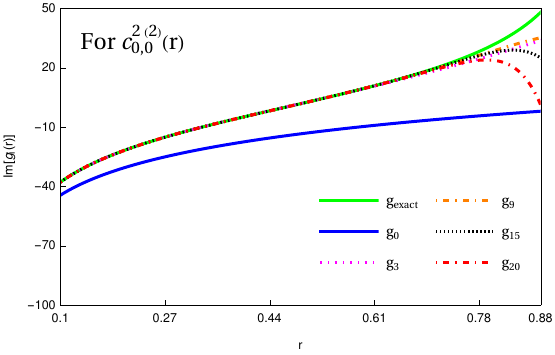}
        \end{subfigure}
        \end{subfigure}
    \end{minipage}
    \caption{Comparison between the imaginary parts of asymptotic expansions around $r=0$ and the exact numerical results.}
    \label{fig:im}
\end{figure}

To examine the convergence behavior of the asymptotic expansion, we truncate the series in $r$ at different orders and compare the resulting approximations with the exact numerical results. The real parts of the SDCs are compared in Fig.~\ref{fig:re}, and the imaginary parts in Fig.~\ref{fig:im}.

We begin by examining the real parts of the SDCs. From Fig.~\ref{fig:re}, several observations can be drawn.
First, the $g_0$ approximation (truncated at $r^0$) exhibits significant deviations from the exact results for most SDCs, even at small $r$. Notable exceptions are $c_{1,0}^{1(2)}$ and $c_{0,1}^{1(2)}$, where $g_0$ provides a good approximation at $r<0.25$. 

Second, as the truncation order increases, the asymptotic expansion converges to the exact values for all the SDCs  at small to moderate $r$. In particular, at $r=0.32$ (corresponding to $m_c=1.5$ GeV and $\sqrt{s}=10.58$ GeV), the asymptotic expansion matches the exact results for all SDCs with relative deviations below $10^{-5}$. This demonstrates that the asymptotic expansion around $r=0$ alone is sufficiently accurate for phenomenological predictions of $J/\psi+\chi_{cJ}$ production at B factories. 

Third, the discrepancy between the asymptotic expansion and the exact results grows sharply near $r=1$,  indicating the presence of singularities at this point. Therefore, a separate expansion around $r=1$ is required to accurately cover the full range $0\le r \le 1$.

Very similar observations can be drawn from Fig.~\ref{fig:im}, which examine the convergence of the imaginary parts of the SDCs. For most SDCs, the asymptotic expansion around $r = 0$ reproduces the exact values well for $r \lesssim 0.7$–$0.8$, depending on the specific SDC. Moreover, including higher-order terms in $r$ further reduces the deviation within this range. It is worth noting that the asymptotic expansions for $c_{1,0}^{0(2)}$ at different truncation orders yield only small discrepancies among themselves, and even the $g_0$ approximation already provides a reasonable estimate for $r < 0.5$.

\subsection{Combined asymptotic expansions around $r=0$ and $r=1$~\label{subsec:r=0:r=1}}

\noindent 
To ensure that the asymptotic expansion accurately reproduces the exact results over the kinematic range $0\le r \le 1$, we perform expansions around both $r=0$ and $r=1$ up to order $20$. Specifically, we take the expansion around $r=0$ for $r<0.45$ and that around $r=1$ for $0.45\le r\le 1$, and compare this combined asymptotic expansion with the exact numerical results.

For convenience, we define the following relative errors:
\begin{subequations}
\begin{align}
    \delta_{\rm{Re}}&=\frac{\rm{Re}[g_{20}]-\rm{Re}[g_{exact}]}{\rm{Re}[g_{exact}]},\\
   \delta_{\rm{Im}}&=\frac{\rm{Im}[g_{20}]-\rm{Im}[g_{exact}]}{\rm{Im}[g_{exact}]}. \label{eq:delta}
\end{align}
\end{subequations}
Note, here $g$ denotes the combined asymptotic expansion (rather than the expansion solely around $r=0$).

\begin{figure} [htbp]
    \centering
    \begin{minipage}{0.48\textwidth}
        \centering
        \begin{subfigure}{\linewidth}
            \includegraphics[width=\linewidth]{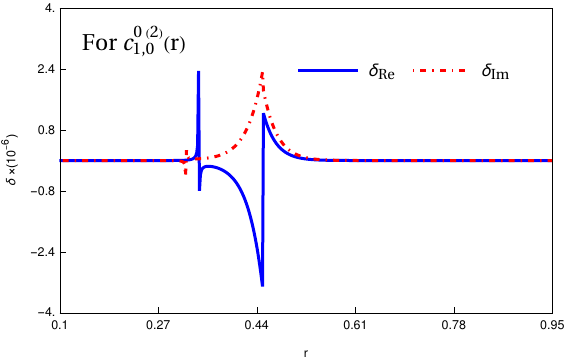}
        \end{subfigure}
        \begin{subfigure}{\linewidth}
            \includegraphics[width=\linewidth]{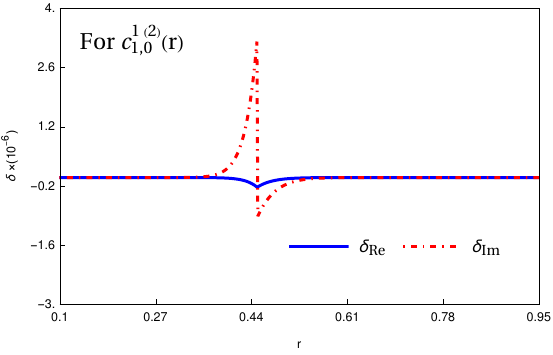}
        \end{subfigure}
        \begin{subfigure}{\linewidth}
            \includegraphics[width=\linewidth]{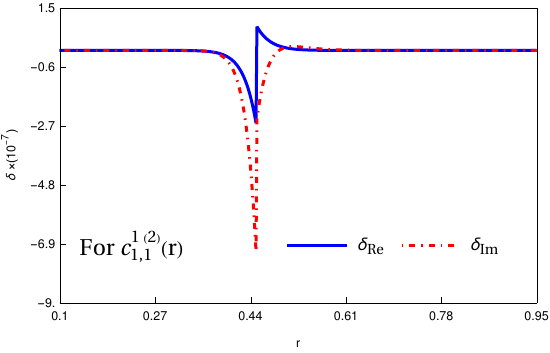}
        \end{subfigure}
        \begin{subfigure}{\linewidth}
            \includegraphics[width=\linewidth]{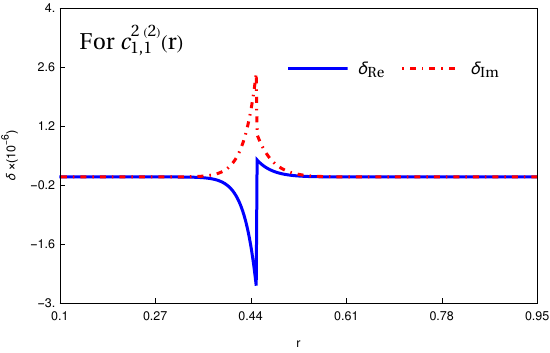}
            \begin{subfigure}{\linewidth}
            \includegraphics[width=\linewidth]{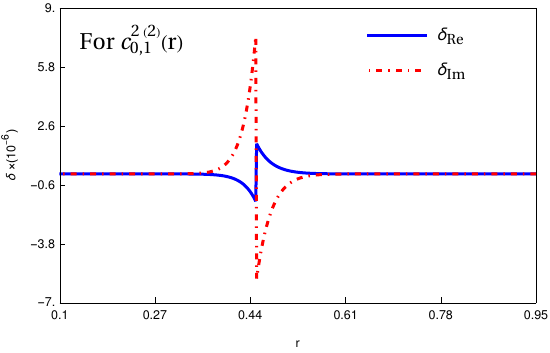}
        \end{subfigure}
        \end{subfigure}
    \end{minipage}
    \hfill
    \begin{minipage}{0.48\textwidth}
        \centering
        \begin{subfigure}{\linewidth}
            \includegraphics[width=\linewidth]{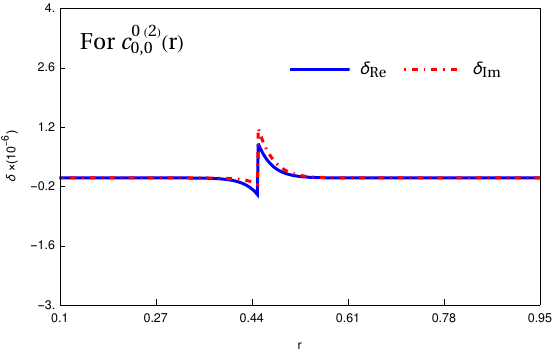}
        \end{subfigure}
        \begin{subfigure}{\linewidth}
            \includegraphics[width=\linewidth]{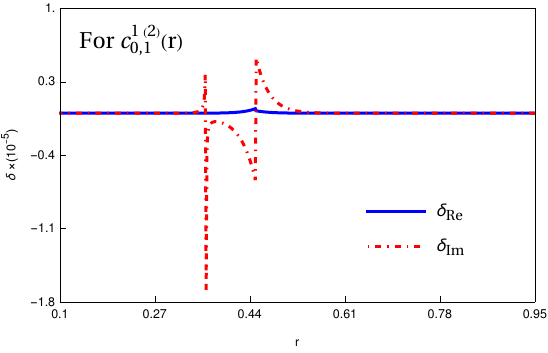}
        \end{subfigure}
        \begin{subfigure}{\linewidth}
            \includegraphics[width=\linewidth]{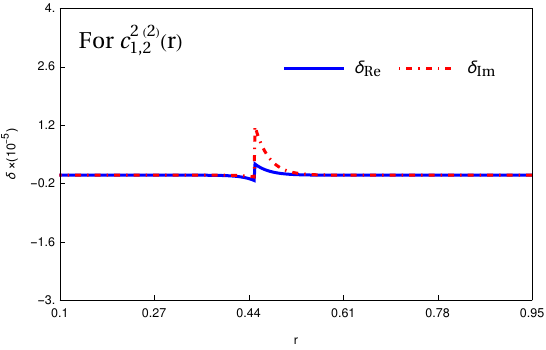}
        \end{subfigure}
        \begin{subfigure}{\linewidth}
            \includegraphics[width=\linewidth]{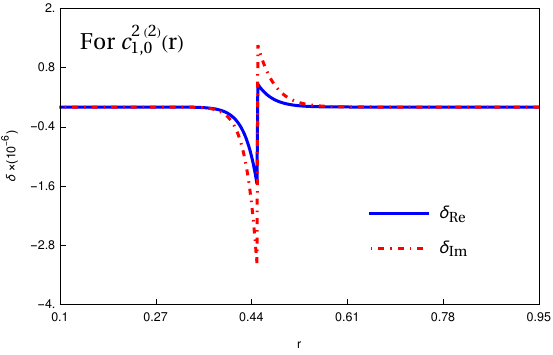}
            \begin{subfigure}{\linewidth}
            \includegraphics[width=\linewidth]{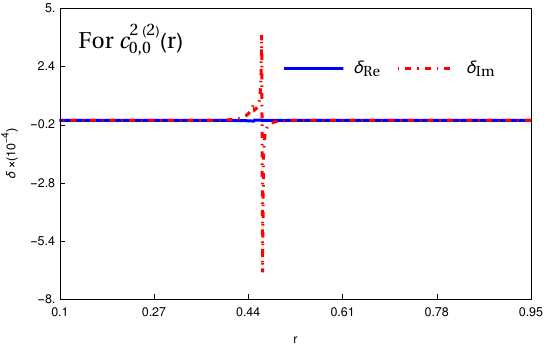}
        \end{subfigure}
        \end{subfigure}
    \end{minipage}
    \caption{Relative errors between the combined asymptotic expansions and the exact results for various SDCs.}
    \label{fig:delta}
\end{figure}

In Fig.~\ref{fig:delta}, we plot $\delta_{\rm{Re}}$ and $\delta_{\rm{Im}}$ as a function of $r$ for various SDCs. The combined asymptotic expansion provides an excellent description over the whole $r$ range, with relative errors well below $10^{-5}$ for most SDCs. One exception is $c^{2\,(2)}_{0,0}$, where the relative error reaches about $10^{-4}$. 

Notably, several peaks are observed in Fig.~\ref{fig:delta}. The peak near $r=0.45$ arises from the piecewise nature of the combined expansion. The peaks to the left of $r=0.45$ are due to the smallness of the absolute values of the exact results in those regions (approaching zero), as clearly seen in Figs.~\ref{fig:re} and \ref{fig:im}. Specifically, small denominators amplify the relative errors. Nevertheless, we have verified that the corresponding absolute errors remain small in these regions.

\section{Phenomenology}\label{sec:pheno}
\subsection{Input parameters}
\noindent 
Prior to making phenomenological predictions, we specify the input parameters used in calculation of the cross sections for $e^+e^-\to J/\psi+\chi_{cJ}$ at B factories. The charm quark mass is taken as $m_c=1.5$ GeV. 

The masses of the quarkonia, which enter through the phase-space integration in Eq.~\eqref{eq:mom} and the normalization of the quarkonium states in Eq.~\eqref{eq:nrqcd}, are set to their physical values as given by the Particle Data Group~\cite{ParticleDataGroup:2024cfk}:
\begin{subequations}
    \begin{align}
        & M_{J/\psi}=3.0969~\mathrm{GeV},\\&
        M_{\chi_{c0}}=3.41471~\mathrm{GeV},\\&
        M_{\chi_{c1}}=3.51067~\mathrm{GeV},\\&
        M_{\chi_{c2}}=3.55617~\mathrm{GeV}.
    \end{align}
\end{subequations}

The CM energy is set to $\sqrt{s}=10.58$ GeV, and the QED coupling constant at this scale is $\alpha=1/130.9$. 
The strong coupling constant $\alpha_s(\mu_R)$ is evaluated with $n_f=4$ active quark flavors. It is obtained by solving the renormalization group equation using the package {\tt RunDec}~\cite{Herren:2017osy,Chetyrkin:2000yt}. To estimate the scale uncertainty, 
we adopt the same convention as in previous studies of exclusive double-charmonium production~\cite{Dong:2011fb,Li:2025mng,Sang:2022kub}, and vary the renormalization scale $\mu_R$ from $2m_c$ to $\sqrt{s}$, with the central value set at $\mu_R=\sqrt{s}/2$.

The NRQCD factorization scale is fixed at $\mu_\Lambda=1$ GeV. The corresponding LDMEs for $J/\psi$ and $\chi_{cJ}$ at this scale are taken from Refs.~\cite{Bodwin:2007fz,Chung:2008km}:
\begin{subequations}
\begin{eqnarray}
\lvert\langle\mathcal{O}_{J/\psi}\rangle\rvert^2&=&0.440^{+0.067}_{-0.055}\,\rm{GeV}^3,\\
\lvert\langle\mathcal{O}_{\chi_{c0}}\rangle\rvert^2&=&0.051\pm 0.010\,\rm{GeV}^5,\\
\lvert\langle\mathcal{O}_{\chi_{c1}}\rangle\rvert^2&=&0.060\pm 0.007\,\rm{GeV}^5,\\
\lvert\langle\mathcal{O}_{\chi_{c2}}\rangle\rvert^2&=&0.068\pm 0.009\,\rm{GeV}^5. 
\end{eqnarray}
\end{subequations}

\subsection{Prediction for the cross sections}
\noindent 
With the input parameters and the derived asymptotic expansions, we calculate the unpolarized cross sections for $e^+e^-\to J/\psi+\chi_{cJ}$ at B factories. The theoretical predictions are summarized in Table~\ref{tab:scacc}, where the experimental results from the {\tt Belle} and {\tt BaBar} collaborations are also shown for comparison. The first theoretical uncertainty quoted in the table originate from the renormalization scale $\mu_R$, while the second one arise from the NRQCD LDMEs.
\begin{table}[htbp]{
    \centering
    \resizebox{\textwidth}{!}{
    \begin{tabular}{|c|c|c|c|c|c|}
    \hline
    & LO & NLO & NNLO & {\tt Belle} \cite{Belle:2004abn} & {\tt BaBar} \cite{BaBar:2005nic}\\
    \hline
      $\sigma(J/\psi+\chi_{c0})$ & $2.66^{+1.03+0.66}_{-0.76-0.62}$ & $5.16^{+1.49+1.28}_{-1.27-1.20}$ & $6.05^{+0.52+1.50}_{-0.96-1.41}$ & $6.4\pm1.7\pm1.0$ & $10.3\pm2.5^{+1.4}_{-1.8}$ \\
      $\sigma(J/\psi+\chi_{c1})$ & $0.54^{+0.21+0.10}_{-0.15-0.09}$ & $0.73^{+0.06+0.14}_{-0.12-0.12}$ & $0.65^{+0.01+0.13}_{-0.12-0.11}$ & -- & --\\
      $\sigma(J/\psi+\chi_{c2})$ & $0.88^{+0.34+0.18}_{-0.25-0.16}$ & $1.10^{+0.04+0.22}_{-0.15-0.20}$ & $0.76^{+0.13+0.15}_{-0.43-0.14}$ & -- & --\\
      $\sigma(J/\psi+\chi_{c1})+\sigma(J/\psi+\chi_{c2})$ & $1.42^{+0.55+0.28}_{-0.41-0.25}$ & $1.82^{+0.10+0.36}_{-0.27-0.32}$ & $1.41^{+0.14+0.28}_{-0.55-0.25}$ & $<5.3$ at $90\%$ C.L. & --\\
      \hline
    \end{tabular}
                 }
    \caption{Cross sections (in fb) for $e^+e^- \to J/\psi + \chi_{cJ}$ at B factories. The first uncertainty originates from varying $\mu_R$ from $2m_c$ to $\sqrt{s}$, with the central value evaluated at $\mu_R=\sqrt{s}/2$ and the second uncertainty arises from the LDMEs.}
    \label{tab:scacc} }
\end{table}

Several observations can be drawn from the table. First, the ${\mathcal{O}}(\alpha_s)$ corrections are positive for all three channels. The correction is significant for $\chi_{c0}$ production, while moderate for $\chi_{c1}$ and $\chi_{c2}$. This explains that the scale uncertainty at NLO becomes larger than that at LO for $\chi_{c0}$,  whereas it is reduced at NLO for $\chi_{c1}$ and $\chi_{c2}$. 

Second, the ${\mathcal{O}}(\alpha_s^2)$ corrections to $\chi_{c0}$ and $\chi_{c1}$ are relatively milder than their corresponding ${\mathcal{O}}(\alpha_s)$ corrections. Specifically for $\chi_{c0}$, the ${\mathcal{O}}(\alpha_s)$ and ${\mathcal{O}}(\alpha_s^2)$ corrections amount to $94\%$ and $33\%$ of the LO cross section, respectively, indicating improved perturbative convergence. In contrast, the ${\mathcal{O}}(\alpha_s^2)$ correction is larger than the ${\mathcal{O}}(\alpha_s)$  correction for $\chi_{c2}$. Notably, the large cancellation between the two corrections for $\chi_{c2}$ brings the NNLO cross section close to the LO prediction.

Third, the QCD corrections reduce the discrepancy between the LO predictions and the experimental data for $J/\psi+\chi_{c0}$. Concretely,  the NNLO prediction agrees with both the {\tt Belle} and {\tt BaBar} data within uncertainties. In addition, the joint cross section for $J/\psi+\chi_{c1}$ and $J/\psi+\chi_{c2}$ is compatible with the upper limit measured by the {\tt Belle} collaboration.

\begin{figure} [htbp]
    \centering
    \begin{subfigure}{0.48\textwidth}
\includegraphics[width=\linewidth]{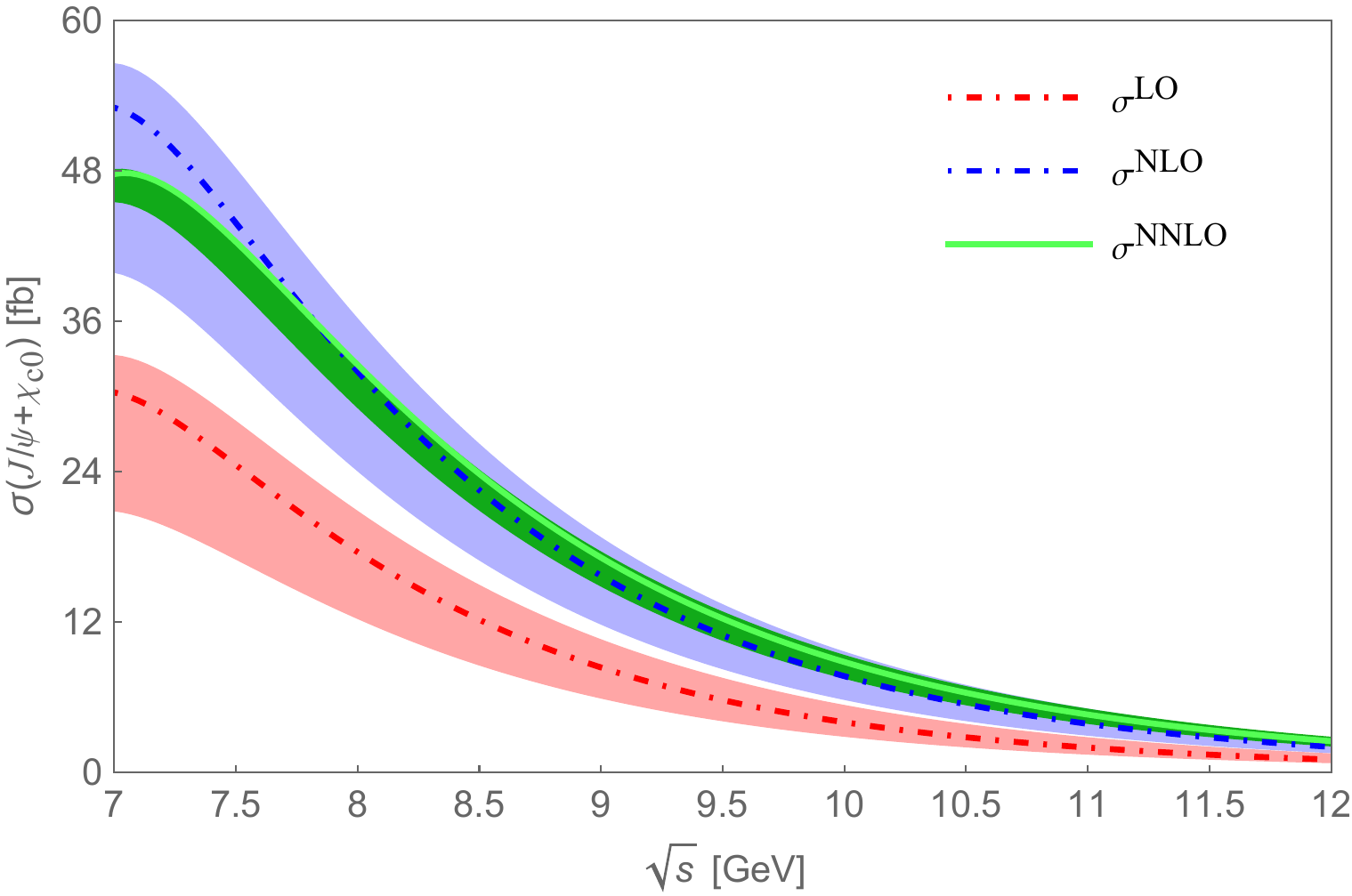}
    \end{subfigure}
    \hfill
    \begin{subfigure}{0.48\textwidth}
\includegraphics[width=\linewidth]{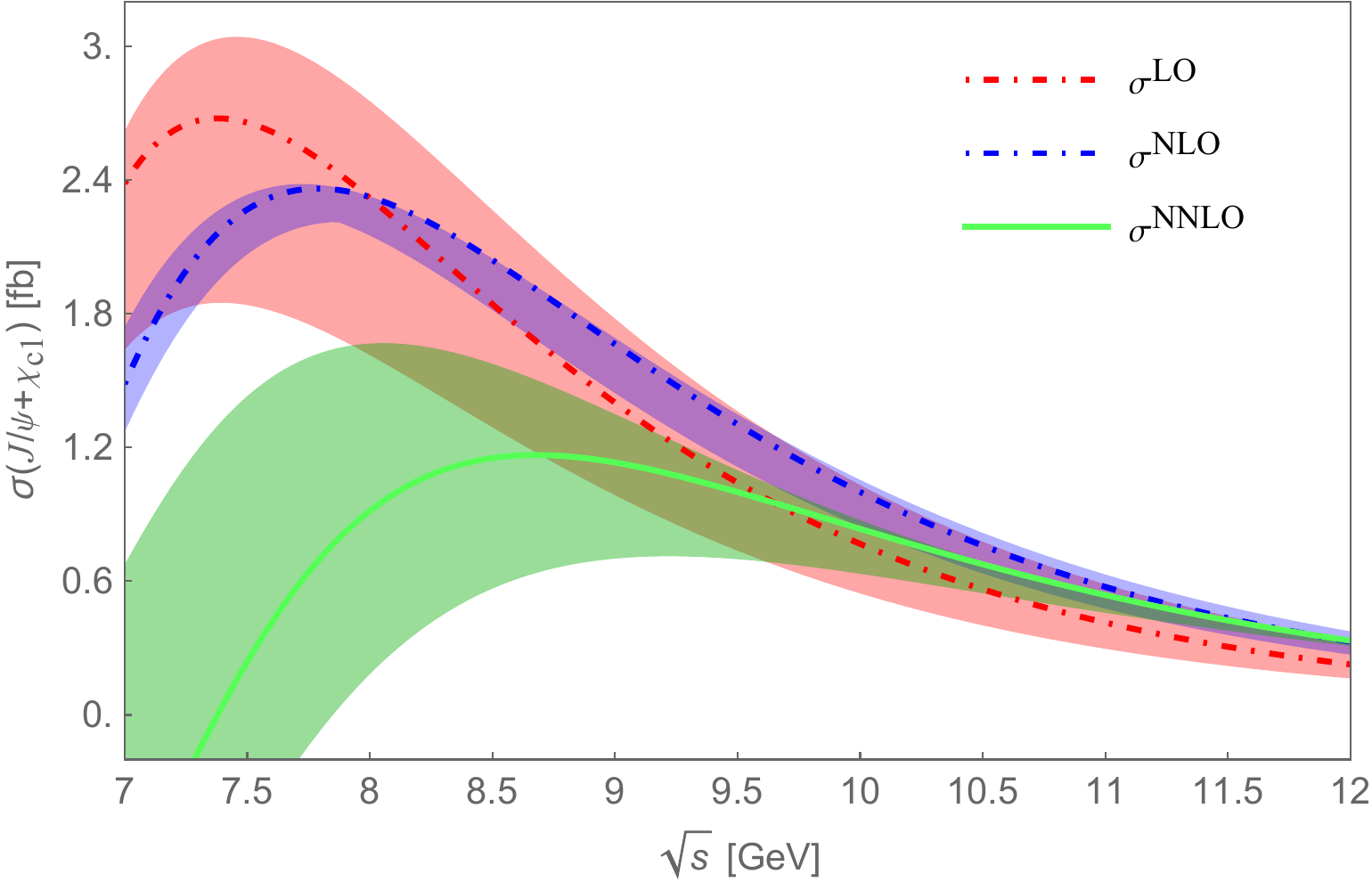}
    \end{subfigure}

    \vspace{0.5em}

    \begin{subfigure}{0.5\textwidth}
\includegraphics[width=\linewidth]{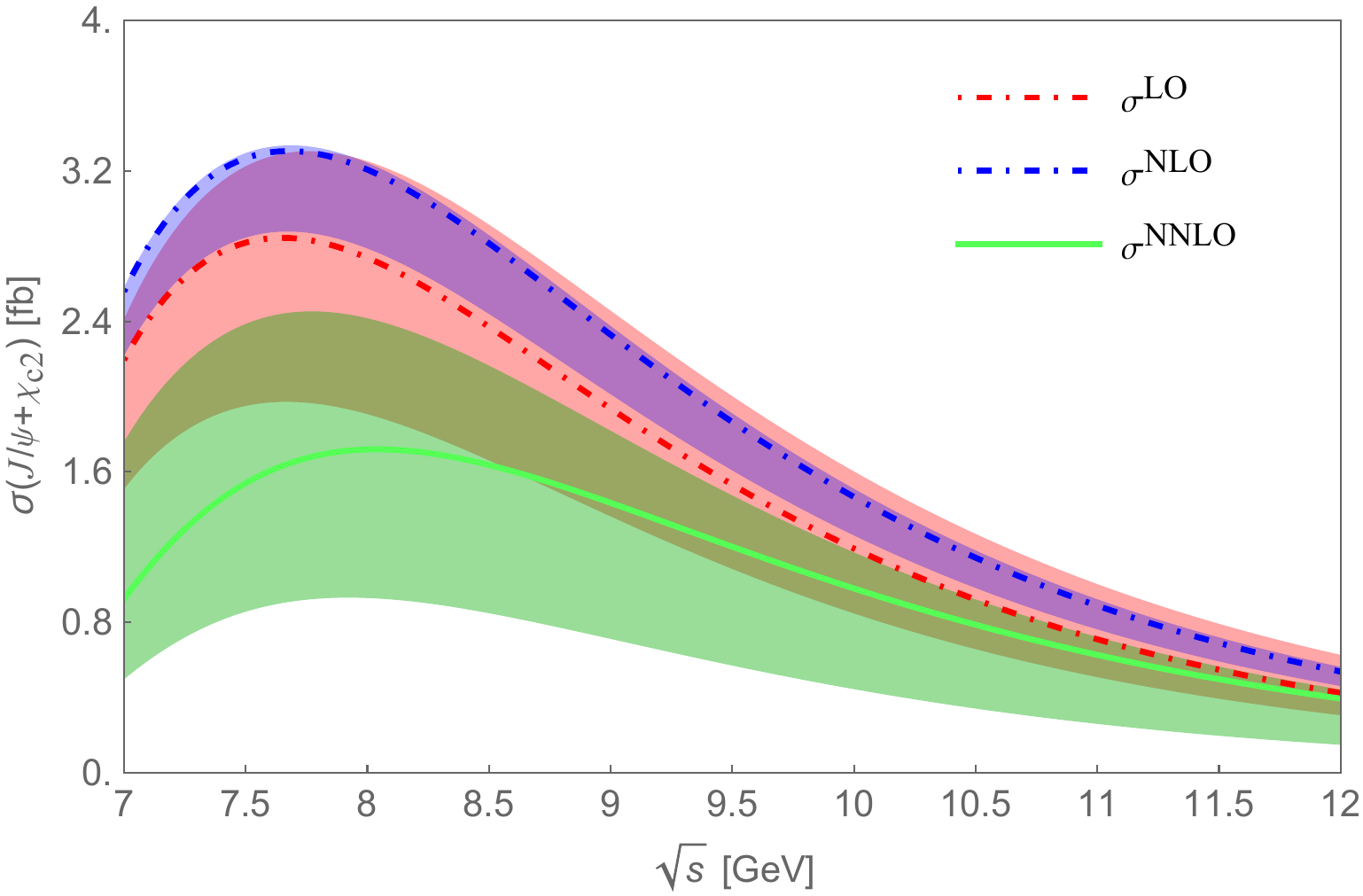}
    \end{subfigure}
    \caption{Cross sections for $e^+e^- \to J/\psi + \chi_{cJ}$  as functions of $\sqrt{s}$ at LO, NLO, and NNLO. The shaded bands represent the renormalization scale uncertainty, obtained by varying $\mu_R$ from $2m_c$ to $\sqrt{s}$, with the central value set to $\mu_R=\sqrt{s}/2$.}
    \label{fig:coresssec}
\end{figure}

Figure~\ref{fig:coresssec} shows the cross sections as functions of $\sqrt{s}$ at different perturbative orders, with uncertainty from scale variation included.~\footnote{The uncertainty from the LDMEs are not included in the figure; they would introduce overall normalization uncertainties of approximately $25\%$, $19\%$, and $20\%$ for the $\chi_{c0}$, $\chi_{c1}$, and $\chi_{c2}$ channels, respectively.} The cross section decreases monotonically for $\chi_{c0}$ production, while it first increases and then decreases for the other two channels. It is observed that the scale uncertainties at NNLO are significantly smaller than those at LO and NLO for $\chi_{c0}$. Although the scale uncertainty for the NNLO prediction of $\chi_{c1}$ is large at lower $\sqrt{s}$, it decreases considerably at higher $\sqrt{s}$. In contrast, the scale uncertainty remains large for the NNLO prediction of $\chi_{c2}$ over the entire considered energy range.   

\subsection{Prediction for the angular distribution parameters $\alpha^J_\theta$}

\noindent 
We present the theoretical predictions for the angular distribution parameters $\alpha^J_\theta$ in Table~\ref{tab:ang}. The values of $\alpha^J_\theta$  at different perturbative orders are obtained by expanding them as a power series in $\alpha_s$. Since $\alpha^J_\theta$ are insensitive to the NRQCD LDMEs, they serve as ideal observables for testing NRQCD predictions. 

Several observations can be drawn from the table. First, the QCD corrections are moderate for $\chi_{c0}$ and $\chi_{c1}$ production. In contrast, the corrections for $\chi_{c2}$ are sizable, indicating poor perturbative convergence for $\alpha^2_\theta$. These large corrections render the value of $\alpha^2_\theta$ at NNLO an order of magnitude smaller than its LO prediction. This behaviour is quite different from that of the cross sections for $\chi_{c2}$, where the NNLO result is close to the LO prediction. We must acknowledge that the instability of the values across different perturbative orders obscures the theoretical reliability.  

Second, in stark contrast to the sizable corrections observed for the cross sections of $\chi_{c0}$ production, the $\mathcal{O}(\alpha_s)$ correction to $\alpha^0_\theta$ is moderate, approximately $14\%$. This is attributed to cancellations between the numerator and denominator in the series expansion of $\alpha^0_\theta$ in $\alpha_s$. It is worth noting that our prediction for $\alpha^0_\theta$ differs significantly from the {\tt Belle} measurement, not only in magnitude but also in sign. The origin of this discrepancy remains unclear. 
Additional experimental determinations of $\alpha^0_\theta$, along with the angular distribution parameters for the $J/\psi+\chi_{c1}$ and $J/\psi+\chi_{c2}$ production channels, would provide valuable tests of our theoretical predictions.

\begin{table}[htbp]
    \centering
    \begin{tabular}{|c|c|c|c|c|}
    \hline
    & LO & NLO & NNLO & {\tt Belle}  \\
    \hline
      $\alpha^0_{\theta}$ & $0.252$ & $0.288^{+0.008}_{-0.006}$ & $0.345^{+0.026}_{-0.017}$ & $-1.01^{+0.38}_{-0.33}$ \\
      $\alpha^1_{\theta}$ & $0.697$ & $0.81^{+0.024}_{-0.018}$ & $0.926^{+0.053}_{-0.036}$ & -- \\
      $\alpha^2_{\theta}$ & $-0.197$ & $-0.105^{+0.019}_{-0.014}$ & $-0.014^{+0.042}_{-0.028}$ & -- \\
      \hline
    \end{tabular}
    \caption{Predictions for the angular distribution $\alpha^J_\theta$. The uncertainties originate from the renormalization scale variation.}
    \label{tab:ang}
\end{table}

\section{Summary}\label{sec:summary}

\noindent 
We present the NNLO QCD corrections to the process $e^+e^-\to J/\psi+\chi_{cJ}$ at B factories within the NRQCD factorization framework. The helicity amplitudes are obtained via asymptotic expansions around both $r=0$ and $r=1$. 
Detailed comparisons show that the asymptotic expressions reproduce the exact numerical results with high accuracy over the kinematic range $0\le r \le 1$, achieving relative errors below $10^{-5}$ for most SDCs, which is sufficient for phenomenological applications. 
Moreover, the large logarithmic terms are obtained analytically using the PSLQ algorithm.  

Using the asymptotic expressions, we compute the unpolarized cross sections. Although the $\mathcal{O}(\alpha_s)$ correction is sizable, the $\mathcal{O}(\alpha_s^2)$ correction is moderate for $\chi_{c0}$ production, leading to a significant reduction of scale uncertainties at NNLO. Both the $\mathcal{O}(\alpha_s)$ and $\mathcal{O}(\alpha_s^2)$ corrections are moderate for $\chi_{c1}$ and $\chi_{c2}$. Interestingly, a large cancellation between the $\mathcal{O}(\alpha_s)$ and $\mathcal{O}(\alpha_s^2)$ corrections brings the NNLO cross section for $\chi_{c2}$  close to the LO prediction. Our prediction for $\chi_{c0}$ is consistent with both the {\tt Belle} and {\tt BaBar} data within uncertainties. In addition, the joint cross section for $J/\psi+\chi_{c1}$ and $J/\psi+\chi_{c2}$ is compatible with the upper limit measured by the {\tt Belle} collaboration.

We also present the angular distribution parameters $\alpha^J_\theta$, which are independent of nonperturbative inputs. 
A sharp discrepancy between theory and {\tt Belle} measurement is observed for $\alpha^0_\theta$, calling for further experimental and theoretical investigations. Moreover, future measurements of the angular distribution parameters for $\chi_{c1}$ and $\chi_{c2}$ will provide important tests of the theoretical framework.

Finally, we emphasize that the asymptotic expansions obtained in this work can be used for further theoretical investigations, such as predictions at different CM energies or with different charm quark masses. These results also shed light on the resummation of large logarithms.

\acknowledgments
The work is supported by the
National Natural Science Foundation of China under Grants
No. 12375079, and Fundamental Research Funds for the Central Universities under Grants No. SWU-KF26003.


\appendix

\section{Construction of helicity projectors} \label{app:hel:pro}
In this appendix, following the method of Refs.~\cite{Dong:2011fb,Xu:2012uh,Zhang:2021ted,Zhang:2022nuf,Sang:2023hjl,Sang:2022erv,Zhang:2023mky}, we construct the ten independent helicity projection operators used in this work.

We first present the transverse metric tensor and the longitudinal vectors, which are required for the construction of the helicity projection operators:
\begin{subequations}
    \begin{align}
 g_{\perp}^{\mu\nu}&=g^{\mu\nu}+\frac{P_1^{\mu}P_1^{\nu}}{\lvert \mathbf{P}\rvert^2}-\frac{Q\cdot P_1}{s\,\lvert \mathbf{P}\rvert^2}(P_1^{\mu}\,Q^{\nu}+Q^{\mu}\,P_1^{\nu})+\frac{M_{J/\psi}^2}{s}\,\frac{Q^{\mu}\,Q^{\nu}}{\lvert \mathbf{P}\rvert^2},\\
         L_{\gamma^*}^{\mu}&=\frac{1}{\lvert \mathbf{P}\rvert}(P_1^{\mu}-\frac{Q\cdot P_1}{s}\,Q^{\mu}),\\
         L_{J/\psi}^{\mu}&=\frac{1}{\lvert \mathbf{P}\rvert}(\frac{Q\cdot P_1}{M_{J/\psi}\,\sqrt{s}}\,P_1^{\mu}-\frac{M_{J/\psi}}{\sqrt{s}}\,Q^{\mu}),\\
         L_{\chi_{cJ}}^{\mu}&=\frac{1}{\lvert \mathbf{P}\rvert}(\frac{Q\cdot P_2}{M_{\chi_{cJ}}\,\sqrt{s}}\,P_2^{\mu}-\frac{M_{\chi_{cJ}}}{\sqrt{s}}\,Q^{\mu})
    \end{align}
\end{subequations}
where $P_1$, $P_2$, and $Q=P_1+P_2$ denote the momenta of the $J/\psi$, the $\chi_{cJ}$, and the virtual photon, respectively. $\lvert \mathbf{P}\rvert$ denotes the magnitude of the three-momentum of the $J/\psi$  in the CM frame, as defined in Eq.~\eqref{eq:mom}. 
Here $P_1^2=M_{J/\psi}^2$, $P_2^2=M_{\chi_{cJ}}^2$, and $Q^2=s$.

For the process $\gamma^*\to J/\psi+\chi_{c0}$, the amplitude $\mathcal{A}^0$ can be expressed as $\mathcal{A}^0=\mathcal{A}^0_{\mu\nu}\,\epsilon^{*\nu}_{J/\psi}(\lambda_1)\,\epsilon^{\mu}_{\gamma^*}$, In this channel, the only two independent helicity combinations that exist are $\lambda_1=1,\lambda_2=0$ and $\lambda_1=0,\lambda_2=0$, whose helicity projection operators are:
\begin{subequations}
    \begin{align}
        \mathcal{P}_{1,0}^{(\chi_{c0})\mu\nu}&=-\frac{1}{2}g_{\perp}^{\mu\nu},\\
        \mathcal{P}_{0,0}^{(\chi_{c0})\mu\nu}&=L_{\gamma^*}^{\mu}\,L_{J/\psi}^{\nu}.
    \end{align}
\end{subequations}
Consequently, the helicity amplitudes are given by $\mathcal{A}^0_{\lambda_1,\lambda_2}=\mathcal{P}_{\lambda_1,0}^{(\chi_{c0})\mu\nu}\,\mathcal{A}^0_{\mu\nu}$ with $\lambda_1=0,1$.

For the process $\gamma^*\to J/\psi+\chi_{c1}$, the amplitude $\mathcal{A}^1$ can be expressed as $\mathcal{A}^1=\mathcal{A}^1_{\mu\nu\alpha}\,\epsilon^{*\nu}_{J/\psi}(\lambda_1)\,\epsilon^{*\alpha}_{\chi_{c1}}(\lambda_2)\,\epsilon^{\mu}_{\gamma^*}$. 
The three independent helicity projection operators are as follows:
\begin{subequations}
    \begin{align}
        \mathcal{P}_{1,1}^{(\chi_{c1})\mu\nu\alpha}&=\frac{i}{2\sqrt{s}\lvert \mathbf{P}\rvert}L_{\gamma^*}^{\mu}\,\epsilon^{\nu\alpha\rho\sigma}\,Q_{\rho}\,P_{1,\sigma},\\
        \mathcal{P}_{1,0}^{(\chi_{c1})\mu\nu\alpha}&=\frac{i}{2\sqrt{s}\lvert \mathbf{P}\rvert}L_{\chi_{c1}}^{\alpha}\,\epsilon^{\mu\nu\rho\sigma}\,Q_{\rho}\,P_{2,\sigma},\\
        \mathcal{P}_{0,1}^{(\chi_{c1})\mu\nu\alpha}&=-\frac{i}{2\sqrt{s}\lvert \mathbf{P}\rvert}L_{J/\psi}^{\nu}\,\epsilon^{\mu\alpha\rho\sigma}\,Q_{\rho}\,P_{1,\sigma}.
    \end{align}
\end{subequations}
The corresponding helicity amplitudes are obtained through $\mathcal{A}^1_{\lambda_1,\lambda_2}=\mathcal{P}_{\lambda_1,\lambda_2}^{(\chi_{c1})\mu\nu\alpha}\,\mathcal{A}^1_{\mu\nu\alpha}$, where $(\lambda_1, \lambda_2)=(1, 1)$, $(1, 0)$ and $(0, 1)$. 

Finally, for the process $\gamma^*\to J/\psi+\chi_{c2}$, the amplitude is $\mathcal{A}^2=\mathcal{A}^2_{\mu\nu\alpha\beta}\,\epsilon^{*\nu}_{J/\psi}(\lambda_1)\,\epsilon^{*\alpha\beta}_{\chi_{c2}}(\lambda_2)\,\epsilon^{\mu}_{\gamma^*}$. The five independent helicity projection operators are:
\begin{subequations}
    \begin{align}
       \mathcal{P}_{1,2}^{(\chi_{c2})\mu\nu\alpha\beta}&=\frac{1}{4}(g_{\perp}^{\mu\nu}\,g_{\perp}^{\alpha\beta}-g_{\perp}^{\mu\alpha}\,g_{\perp}^{\nu\beta}-g_{\perp}^{\mu\beta}\,g_{\perp}^{\nu\alpha}),\\
       \mathcal{P}_{1,1}^{(\chi_{c2})\mu\nu\alpha\beta}&=-\frac{1}{2\sqrt{2}}L_{\gamma^*}^{\mu}(g_{\perp}^{\nu\alpha}\,L_{\chi_{cJ}}^{\beta}+g_{\perp}^{\nu\beta}\,L_{\chi_{cJ}}^{\alpha}),\\
       \mathcal{P}_{1,0}^{(\chi_{c2})\mu\nu\alpha\beta}&=-\frac{1}{2\sqrt{6}}g_{\perp}^{\mu\nu}(g_{\perp}^{\alpha\beta}+2L_{\chi_{cJ}}^{\alpha}\,L_{\chi_{cJ}}^{\beta}),\\
       \mathcal{P}_{0,1}^{(\chi_{c2})\mu\nu\alpha\beta}&=\frac{1}{2\sqrt{2}}L_{J/\psi}^{\nu}(g_{\perp}^{\mu\alpha}\,L_{\chi_{cJ}}^{\beta}+g_{\perp}^{\mu\beta}\,L_{\chi_{cJ}}^{\alpha}),\\
     \mathcal{P}_{0,0}^{(\chi_{c2})\mu\nu\alpha\beta}&=\frac{1}{\sqrt{6}}L_{\gamma^*}^{\mu}\,L_{J/\psi}^{\nu}(g_{\perp}^{\alpha\beta}+2L_{\chi_{cJ}}^{\alpha}\,L_{\chi_{cJ}}^{\beta}).
    \end{align}
\end{subequations}
By applying these projection operators, the helicity amplitudes are computed via $\mathcal{A}^2_{\lambda_1,\lambda_2}=\mathcal{P}_{\lambda_1,\lambda_2}^{(\chi_{c2})\mu\nu\alpha\beta}\,\mathcal{A}^2_{\mu\nu\alpha\beta}$, where $(\lambda_1,\lambda_2)=(1, 2)$, $(1, 1)$, $(1, 0)$, $(0, 1)$ and $(0, 0)$. 

It is worth noting that since we neglect relativistic corrections in the computation of the helicity amplitudes, we adopt the approximations $M_{J/\psi}\approx 2m_c$, $M_{\chi_{cJ}}\approx 2m_c$. 

\section{Asymptotic expression at $r=0$}\label{app:asy:exp}
\noindent 

In this appendix, we present the expansion coefficients for the SDCs up to $\mathcal{O}(r^0)$. Specifically, the $\mathcal{O}(\alpha_s)$ coefficients $f^{J\,(1)}_{\lambda_1,\lambda_2}$ are given in appendix~\ref{app1}, and the $\mathcal{O}(\alpha_s^2)$ coefficients $f^{J\,(2)}_{\lambda_1,\lambda_2}$ are given in appendix~\ref{app2}. The definition of these expansion coefficients follows Eq.~\eqref{eq:ff}.

Here, the subscript $lbl$ denotes contributions from the ``light-by-light" Feynman diagrams, as illustrated in Fig.~\ref{fig:feynmanpic}d. In this appendix, we retain the full color factors for the $\mathcal{O}(\alpha_s)$ coefficients $f^{J\,(1)}_{\lambda_1,\lambda_2}$. For the $\mathcal{O}(\alpha_s^2)$ coefficients, only partial color factors are kept due to space limitations. Complete expansion coefficients including the full color-factor dependence, up to $\mathcal{O}(r^{20})$, are provided in the attached electronic supplementary material files.

\subsection{$\mathcal{O}(\alpha_s)$ expanding coefficients up to $\mathcal{O}(r^0)$~\label{app1}}
\begin{flalign}
&\begin{aligned}
f^{0\,(1)}_{1,0}(0,2) &= \frac{27}{8}C_F-\frac{1}{2}C_A , f^{0\,(1)}_{0,0}(0,2) = 0 ,\notag
\end{aligned}&\\ \notag
&\begin{aligned}
&f^{1\,(1)}_{1,0}(0,2) = \sqrt{6}\bigg(\frac{7}{8}C_F-\frac{1}{4}C_A\bigg) ,\\& f^{1\,(1)}_{0,1}(0,2) = \sqrt{6}\bigg(\frac{1}{8}C_A-\frac{17}{16}C_F\bigg),f^{1\,(1)}_{1,1}(0,2) = -\frac{5}{8}\sqrt{6}C_F,
\end{aligned}&\\ \notag
&\begin{aligned} 
&f^{2\,(1)}_{1,2}(0,2) = \frac{\sqrt{3}}{2}\bigg(C_A-3C_F+6T_F\bigg) , f^{2\,(1)}_{1,1}(0,2) = \sqrt{6}\bigg(\frac{C_A}{2}-\frac{11}{8}C_F\bigg),\\&f^{2\,(1)}_{1,0}(0,2) = \frac{\sqrt{2}}{2}\bigg(C_A-\frac{15}{4}C_F\bigg), f^{2\,(1)}_{0,1}(0,2) =\sqrt{6}\bigg(\frac{1}{8}C_A-\frac{11}{16}C_F\bigg),\\&f^{2\,(1)}_{0,0}(0,2) = 0, 
\end{aligned}&\\ 
\end{flalign}
\begin{flalign}
&\begin{aligned}
f^{0\,(1)}_{1,0}(0,1) &=(-\frac{3}{2}-i\pi)C_A+(-\frac{41}{8}-\frac{7\ln2}{4}+i\frac{27\pi}{4})C_F+(-\frac{1}{2}-2\ln2)T_F, \notag
\end{aligned}&\\ \notag
&\begin{aligned}
f^{0\,(1)}_{0,0}(0,1) &=(\ln2-1)C_F-2\ln2\;T_F,
\end{aligned}&\\ \notag
&\begin{aligned}
f^{1}_{1,0}(0,1) &=\frac{1}{\sqrt{6}}\bigg[(-3 \ln2-3 i\pi)C_A+(\frac{15}{4}-\frac{3 \ln2}{2}+\frac{21 i\pi }{2})C_F-3\;T_F\bigg],
\end{aligned}&\\ \notag
&\begin{aligned}
f^{1\,(1)}_{0,1}(0,1) &=\frac{1}{\sqrt{6}}\bigg[(3-\frac{3 \ln2}{2}+\frac{3i \pi }{2})C_A+(\frac{33}{4}-\frac{21 \ln2}{2} -\frac{51 i\pi }{4})C_F+(6-12 \ln2)T_F\bigg],
\end{aligned}&\\ \notag
&\begin{aligned}
f^{1\,(1)}_{1,1}(0,1) &=\frac{1}{\sqrt{6}}\bigg[(3-3 \ln2)C_A+(9 \ln2-\frac{15}{2}-\frac{15 i\pi }{2})C_F\bigg],
\end{aligned}&\\ \notag
&\begin{aligned}
f^{2\,(1)}_{1,2}(0,1) &= \frac{1}{\sqrt{3}}\bigg[(21 \ln2-3+3 i\pi)C_A+(\frac{21}{4}-\frac{75 \ln2}{2}-9 i\pi)C_F+(6-24 \ln2+6 i\pi)T_F\bigg],
\end{aligned}&\\ \notag
&\begin{aligned}
f^{2\,(1)}_{1,1}(0,1) &= \frac{1}{\sqrt{6}}\bigg[(15-9 \ln2+6 i\pi)C_A+(3 \ln2-\frac{33}{2}-\frac{33i \pi }{2})C_F\bigg],
\end{aligned}&\\ \notag
&\begin{aligned}
f^{2\,(1)}_{1,0}(0,1) &= \frac{1}{\sqrt{2}}\bigg[(3+2i\pi)C_A+(\frac{7 \ln2}{2}-\frac{31}{4}-\frac{15 i\pi }{2})C_F+(1+4 \ln2)T_F\bigg],
\end{aligned}&\\ \notag
&\begin{aligned}
f^{2\,(1)}_{0,1}(0,1) &= \frac{1}{\sqrt{6}}\bigg[(3-\frac{3 \ln2}{2}+\frac{3 i\pi }{2})C_A+(\frac{3 \ln2}{2}-\frac{21}{4}-\frac{33 i\pi }{4})C_F+(18-12 \ln2)T_F\bigg],
\end{aligned}&\\ \notag
&\begin{aligned}
f^{2\,(1)}_{0,0}(0,1) &= (\sqrt{2}-\sqrt{2}\ln2)C_F+2\sqrt{2}\ln2\;T_F,
\end{aligned}&\\
\end{flalign}
\begin{flalign}
&\begin{aligned}
   \notag  f^{0\,(1)}_{1,0}(0,0) &= (\frac{241}{12}+\frac{\pi ^2}{6}+\ln^22+\frac{13 \ln2}{2}+\frac{27 i\pi }{4})C_A+\big[-\frac{235}{8}-\frac{5 \pi ^2}{6}-\frac{159 \ln^22}{8}\\&+\frac{267 \ln2}{8}+i(-\frac{41 \pi }{8}-\frac{7}{4} \pi  \ln2)\big]C_F+\big[-\frac{17 n_h}{3}-\frac{17 n_l}{3}\\&+i(-3 \pi  n_h-3 \pi  n_l)-\frac{7}{3}+\frac{\pi ^2}{3}+3 \ln^22+\frac{5 \ln2}{6}+i(-\frac{\pi }{2}-2\pi  \ln2)\big]T_F,
\end{aligned}&\\ \notag
&\begin{aligned}
f^{0\,(1)}_{0,0}(0,0) &=(\frac{28}{9}-\frac{4 \ln2}{3}+\frac{11 i\pi }{12})C_A+\big[ -\frac{19}{6}-\frac{\pi ^2}{6}-3 \ln^22+\frac{43 \ln2}{6}\\&+i( \pi  \ln2-\pi)\big]C_F+\big[-\frac{8 n_h}{9}-\frac{8 n_l}{9}+i( -\frac{\pi  n_h}{3}-\frac{\pi  n_l}{3})\\&-\frac{2}{3}+\frac{\pi ^2}{3}+3 \ln^22-\frac{4 \ln2}{3}-2i\pi  \ln2\big]T_F,
\end{aligned}&\\ \notag
&\begin{aligned}
f^{1\,(1)}_{1,0}(0,0) &= \frac{1}{\sqrt{6}}\bigg\{(-\frac{3}{2}+\pi ^2+\frac{15 \ln^22}{2}-3 \ln2-3 i\pi  \ln2)C_A\\&+\big[-\frac{15}{4}-\frac{3 \pi ^2}{2}-\frac{63 \ln^22}{4}+\frac{87 \ln2}{4}+i(\frac{15 \pi }{4}-\frac{3}{2} \pi  \ln2)\big]C_F\\&+(21 \ln2-3i \pi)T_F\bigg\},
\end{aligned}&\\ \notag
&\begin{aligned}
f^{1\,(1)}_{0,1}(0,0) &= \frac{1}{\sqrt{6}}\bigg\{\big[ -\frac{76}{3}+\frac{15 \ln^22}{4}-22 \ln2+i(-8 \pi -\frac{3}{2} \pi  \ln2)\big]C_A\\&+\big[\frac{335}{8}+\frac{31 \pi ^2}{8}+\frac{237 \ln^22}{4}-82 \ln2+i(\frac{33 \pi }{4}-\frac{21}{2} \pi  \ln2)\big]C_F\\&+\big[\frac{26 n_h}{3}+\frac{26 n_l}{3}+i(4 \pi  n_h+4 \pi  n_l)+2\pi ^2+18 \ln^22-24 \ln2\\&+i(6 \pi -12 \pi  \ln2)\big]T_F\bigg\},
\end{aligned}&\\ \notag
&\begin{aligned}
f^{1\,(1)}_{1,1}(0,0) &=\frac{1}{\sqrt{6}}\bigg\{\big[-\frac{161}{6}+\frac{\pi ^2}{2}+\frac{3 \ln^22}{2}-10 \ln2+i( -8 \pi -3 \pi  \ln2)\big]C_A\\&+\big[40-\frac{\pi ^2}{4}+\frac{51 \ln^22}{2}-\frac{11 \ln2}{2}+i(9 \pi  \ln2-\frac{15 \pi }{2})\big]C_F\\&+\big[\frac{26 n_h}{3}+\frac{26 n_l}{3}+i(4 \pi  n_h+4 \pi  n_l)+36 \ln2\big]T_F\bigg\},
\end{aligned}&\\ \notag
&\begin{aligned} \notag
f^{2\,(1)}_{1,2}(0,0) &=\frac{1}{\sqrt{3}}\bigg\{\big[ -\frac{97}{6}-4 \pi ^2-\frac{57 \ln^22}{2}+50 \ln2+i(21 \pi  \ln2-\frac{17 \pi }{2})\big]C_A\\&+\big[\frac{127}{4}+\frac{31 \pi ^2}{4}+\frac{333 \ln^22}{4}-\frac{493 \ln2}{4}+i(\frac{21 \pi }{4}-\frac{75}{2} \pi  \ln2)\big]C_F\\&+\big[\frac{10 n_h}{3}+\frac{10 n_l}{3}+i(2 \pi  n_h+2 \pi  n_l)+4+\pi ^2-10 \ln2-5 i\pi \big]T_F\bigg\},
\end{aligned}&\\ \notag
&\begin{aligned}
f^{2\,(1)}_{1,1}(0,0) &=\frac{1}{\sqrt{6}}\bigg\{\big[-\frac{101}{6}+\frac{\pi ^2}{2}+\frac{33 \ln^22}{2}-68 \ln2+i(4 \pi -9 \pi  \ln2)\big]C_A\\&+\big[20+\frac{9 \pi ^2}{4}+\frac{9 \ln^22}{2}+\frac{155 \ln2}{2}+i(3 \pi  \ln2-\frac{33 \pi }{2})\big]C_F\\&+\big[\frac{26 n_h}{3}+\frac{26 n_l}{3}+i( 4 \pi  n_h+4 \pi  n_l)+4 \ln2-4\big]T_F\bigg\},
\end{aligned}&\\ \notag
&\begin{aligned}
f^{2\,(1)}_{1,0}(0,0) &=\frac{1}{\sqrt{2}}\bigg\{(-\frac{49}{3}-\frac{\pi ^2}{3}-2 \ln^22-\ln2-\frac{5 i\pi }{2})C_A\\&+\big[\frac{79}{4}+\frac{2 \pi ^2}{3}+\frac{63 \ln^22}{4}+\frac{21 \ln2}{4}+i(\frac{7}{2} \pi  \ln2-\frac{31 \pi }{4})\big]C_F\\&+\big[\frac{14 n_h}{3}+\frac{14 n_l}{3}+i(2 \pi  n_h+2 \pi  n_l)+\frac{2}{3}-\frac{2\pi ^2}{3}-6 \ln^22+\frac{43 \ln2}{3}\\&+i(4 \pi  \ln2-\pi )\big]T_F\bigg\}, 
\end{aligned}&\\ \notag
&\begin{aligned}
f^{2\,(1)}_{0,1}(0,0) &=\frac{1}{\sqrt{6}}\bigg\{\big[-\frac{97}{6}+\frac{15 \ln^22}{4}+\ln2+i(-\frac{5 \pi }{2}-\frac{3}{2} \pi  \ln2)\big]C_A\\&+\big[\frac{139}{8}+\frac{9 \pi ^2}{8}+\frac{69 \ln^22}{4}-23 \ln2+i(\frac{3}{2} \pi  \ln2-\frac{21 \pi }{4})\big]C_F\\&+\big[\frac{16 n_h}{3}+\frac{16 n_l}{3}+i(2 \pi  n_h+2 \pi  n_l)+10+2\pi ^2+18 \ln^22-52 \ln2\\&+i(10 \pi -12 \pi  \ln2)\big]T_F\bigg\},
\end{aligned}&\\ \notag
&\begin{aligned}
f^{2\,(1)}_{0,0}(0,0) &=\frac{1}{\sqrt{2}}\bigg\{(\frac{8 \ln2}{3}-\frac{56}{9}-\frac{11 i\pi }{6})C_A+\big[\frac{28}{3}+\frac{\pi ^2}{3}+6 \ln^22\\&-\frac{25 \ln2}{3}+i(2 \pi -2 \pi  \ln2 )\big]C_F+\big[\frac{16 n_h}{9}+\frac{16 n_l}{9}\\&+i(\frac{2 \pi  n_h}{3}+\frac{2 \pi  n_l}{3})-\frac{8}{3}-\frac{2\pi ^2}{3}-6 \ln^22+\frac{56 \ln2}{3}\\&+i(4 \pi  \ln2-2\pi)\big]T_F\bigg\}.
\end{aligned}&\\
\end{flalign}
\subsection{$\mathcal{O}(\alpha_s^2)$ expanding coefficients up to $\mathcal{O}(r^0)$~\label{app2}}
\begin{flalign}
&\begin{aligned} \notag
f^{2\,(2)}_{1,2}(-1,2)&=\frac{1}{\sqrt{3}}\bigg[3C_AT_F+(-\frac{9}{2}-3\ln2)C_FT_F\bigg],
\end{aligned}\\& \notag
\begin{aligned}
\notag f^{2\,(2)}_{1,2}(-1,1)&=\frac{1}{\sqrt{3}}\bigg\{(-6-18 \ln2+6 i\pi)C_AT_F+\big[6+\pi ^2+21 \ln^22+21 \ln2\\&+i(-9 \pi -6 \pi  \ln2)\big]C_FT_F\bigg\},
\end{aligned}\\&
\begin{aligned} \notag
f^{2\,(2)}_{1,2}(-1,0)&=\frac{1}{\sqrt{3}}\bigg\{\big[6-2\pi ^2+27 \ln^22+i(-6 \pi -18 \pi  \ln2)\big]C_AT_F\\&+\big[-12 \zeta (3)-6+4 \pi ^2-37 \ln^32+\frac{9 \ln^22}{2}-2\pi ^2 \ln2\\&+i(6 \pi +\pi ^3+21 \pi  \ln^22+21 \pi  \ln2)\big]C_FT_F\bigg\},
\end{aligned}&\\
\end{flalign}
\begin{flalign}
&\begin{aligned}
&f^{0\,(2)}_{1,0}(0,4) =\frac{13 C_F^2}{64}-\frac{5 C_A C_F}{96},f^{0\,(2)}_{0,0}(0,4) =0,\\&f^{1\,(2)}_{1,0}(0,4) =\sqrt{6}\bigg(\frac{5}{64}  C_F^2-\frac{1}{32}  C_A C_F\bigg),\notag
\end{aligned}&\\ \notag
&\begin{aligned}
f^{1\,(2)}_{0,1}(0,4) &=\frac{1}{ \sqrt{6}}\bigg(\frac{C_A C_F}{8}-\frac{29 C_F^2}{64}\bigg),f^{1\,(2)}_{1,1}(0,4) =\frac{1}{\sqrt{6}}\bigg(\frac{C_A C_F}{8 }-\frac{13 C_F^2}{32 }\bigg)\notag
\end{aligned}&\\ \notag
&\begin{aligned}
f^{2\,(2)}_{1,2}(0,4) &=\frac{1}{\sqrt{3}}\bigg(-\frac{7 C_AT_F}{16 }-\frac{C_F^2}{16 }-\frac{3 C_FT_F}{16}\bigg),f^{2\,(2)}_{1,1}(0,4) =\frac{1}{\sqrt{6}}\bigg(\frac{C_A C_F}{8 }-\frac{11 C_F^2}{32 }\bigg),
\end{aligned}&\\ \notag
&\begin{aligned}
f^{2\,(2)}_{1,0}(0,4) &=\frac{1}{\sqrt{2}}\bigg(\frac{5 C_A C_F}{48 }-\frac{9 C_F^2}{32 }\bigg),f^{2\,(2)}_{0,1}(0,4) =\frac{1}{\sqrt{6}}\bigg(\frac{C_A C_F}{8 }-\frac{23 C_F^2}{64 }\bigg),f^{2\,(2)}_{0,0}(0,4) = 0,
\end{aligned}&\\
\end{flalign}
\begin{flalign}
&\begin{aligned}
f^{0\,(2)}_{1,0}(0,3) &=( \frac{29}{144}-\frac{\ln2}{24} )C_A^2+\big[-\frac{7}{4}-\frac{\ln2}{4}-\frac{5 i\pi }{24}\big]C_AC_F\\&+(\frac{49 \ln2}{24}-\frac{29}{12}+\frac{13 i\pi }{16})C_F^2+(-\frac{17}{24}-\frac{\ln2}{2})C_FT_F-\frac{1}{36} (n_h+n_l) \left(4 C_A-27 C_F\right)T_F, \notag
\end{aligned}&\\ \notag
&\begin{aligned}
f^{0\,(2)}_{0,0}(0,3) &=0,
\end{aligned}&\\ \notag
&\begin{aligned}
f^{1\,(2)}_{1,0}(0,3) &=\frac{1}{\sqrt{6}}\bigg\{( \frac{1}{4}-\ln2 )C_AT_F+(\frac{2}{3}-\frac{\ln2}{8})C_A^2+(-\frac{53}{24}-\ln2 -\frac{3i \pi }{4})C_AC_F\\&+\big[\frac{1}{4} (17 \ln2-15)+\frac{15 i\pi }{8}\big]C_F^2+(2\ln2-\frac{3}{4})C_FT_F-\frac{1}{6} (n_h+n_l) \left(2 C_A-7 C_F\right)T_F\bigg\},
\end{aligned}&\\ \notag
&\begin{aligned}
f^{1\,(2)}_{0,1}(0,3) &=\frac{1}{\sqrt{6}}\bigg\{\frac{1}{48} (3\ln2-22)C_A^2+( \frac{229}{48}-\frac{3 \ln2}{8}+\frac{i\pi }{2})C_AC_F\\&+\big[\frac{5}{8} (7-10 \ln2) -\frac{29i \pi }{16}\big]C_F^2+\frac{1}{12} (n_h+n_l) \left(2 C_A-17 C_F\right)T_F\bigg\},
\end{aligned}&\\ \notag
&\begin{aligned}
f^{1\,(2)}_{1,1}(0,3) &=\frac{1}{\sqrt{6}}\bigg[(2\ln2-1)C_AT_F-\frac{3}{8} (\ln2-1)C_A^2+(\frac{67}{24}-\frac{\ln2}{4}+ \frac{i\pi }{2})C_AC_F\\&+(-\frac{11}{8}-\frac{\ln2}{2}-\frac{13i \pi }{8})C_F^2+\big[3-6\ln2-\frac{5}{6}  (n_h+n_l)\big]C_F T_F\bigg],
\end{aligned}&\\ \notag
&\begin{aligned}
f^{2\,(2)}_{1,2}(0,3) &=\frac{1}{\sqrt{3}}\bigg\{( 3 \ln2-\frac{127}{12})C_AT_F+(\frac{3 \ln2}{8}-\frac{23}{12})C_A^2+\frac{9}{8} (5+3\ln2)C_AC_F\\&+(-\frac{9}{16}-\frac{17 \ln2}{2} -\frac{i\pi }{4})C_F^2+ (17 \ln2-\frac{83}{8}-n_l-n_h)C_FT_F\\&+\frac{1}{3}(n_l+n_h)C_AT_F+\frac{1}{3}(17n_l+13n_h)T_F^2\bigg\},
\end{aligned}&\\ \notag
&\begin{aligned}
f^{2\,(2)}_{1,1}(0,3) &=\frac{1}{\sqrt{6}}\bigg\{-C_AT_F+\frac{7}{24} (3\ln2-5)C_A^2+( \frac{109}{24}-\frac{3 \ln2}{4}+\frac{i\pi }{2})C_AC_F\\&+\big[\frac{1}{8} (13-44 \ln2)-\frac{11 i\pi }{8}\big]C_F^2+(4-4\ln2)C_FT_F+\frac{1}{6} (n_h+n_l) \left(4 C_A-11 C_F\right)T_F\bigg\},
\end{aligned}&\\ \notag
&\begin{aligned}
f^{2\,(2)}_{1,0}(0,3) &=\frac{1}{\sqrt{2}}\bigg\{-\frac{5}{12}C_AT_F+\frac{1}{72} (6\ln2-29) C_A^2+\big[\frac{1}{6} (13+3\ln2)+ \frac{5 i\pi }{12}\big]C_AC_F\\&+\big[\frac{1}{12} (4-25 \ln2)-\frac{9 i\pi }{8}\big]C_F^2+(\frac{11}{12}+\ln2)C_FT_F+\frac{1}{18} (n_h+n_l) \left(4 C_A-15 C_F\right)T_F\bigg\},
\end{aligned}&\\ \notag
&\begin{aligned}
f^{2\,(2)}_{0,1}(0,3) &=\frac{1}{\sqrt{6}}\bigg[ \frac{1}{48} (3\ln2-22)C_A^2+( \frac{169}{48}-\frac{3 \ln2}{8}+\frac{i\pi }{2})C_AC_F+(\frac{3}{8}-\frac{9 \ln2}{4} -\frac{23i \pi }{16})C_F^2\\&+\frac{3}{2}C_FT_F+\frac{1}{12} (n_h+n_l) \left(2 C_A-11 C_F\right)T_F\bigg],
\end{aligned}&\\ \notag
&\begin{aligned}
f^{2\,(2)}_{0,0}(0,3) &=0,
\end{aligned}&\\
\end{flalign}
\begin{flalign}
&\begin{aligned}
f^{0\,(2)}_{1,0}(0,2) &=\big[(\frac{5 \ln^22}{24}-\frac{5 \ln2}{48})_{lbl}-\frac{109}{36}-\frac{49 \ln2}{36}\big]n_h+(-\frac{109}{36}-\frac{49 \ln2}{36})n_l\\&+\frac{6607}{144}-\frac{1621 \pi ^2}{1728}-\frac{4889 \ln^22}{288}+\frac{6073 \ln2}{144}+i(\frac{415}{72} \pi  \ln2-\frac{1709 \pi }{144}),\notag
\end{aligned}&\\ \notag
&\begin{aligned}
f^{0\,(2)}_{0,0}(0,2) &=(-\frac{1}{4}+\frac{\pi ^2}{8}+\frac{\ln^22}{2}-\frac{2\ln2}{3})C_AT_F+(\frac{11}{24}-\frac{11 \ln2}{24})C_AC_F\\& \notag+(\frac{1}{4}-\frac{\pi ^2}{48}+\frac{3 \ln^22}{8}-\frac{7 \ln2}{8})C_F^2+\big[\frac{1}{2}-\frac{\pi ^2}{24}-\frac{3 \ln^22}{2}+\frac{3 \ln2}{2}\\&+(\frac{\ln2}{6}-\frac{1}{6})n_h+(\frac{\ln2}{6}-\frac{1}{6})n_l\big]C_FT_F+(-\frac{2}{3} \ln2\,n_h-\frac{2}{3} \ln2\,n_l)T_F^2,
\end{aligned}&\\ \notag
&\begin{aligned}
f^{1\,(2)}_{1,0}(0,2) &=\frac{1}{\sqrt{6}}\bigg\{\big[(\frac{5 \ln^22}{8}-\frac{5}{16})_{lbl}-\frac{83}{144}-\frac{7 \ln2}{4}\big]n_h+(-\frac{83}{144}-\frac{7 \ln2}{4})n_l-\frac{463}{48}\\&-\frac{13 \pi ^2}{72}-\frac{133 \ln^22}{8}+64 \ln2+i(\frac{163}{24} \pi  \ln2-\frac{121 \pi }{8})\bigg\},
\end{aligned}&\\ \notag
&\begin{aligned}
f^{1\,(2)}_{0,1}(0,2) &=\frac{1}{\sqrt{6}}\bigg\{\big[(\frac{5 \ln2}{16}-\frac{5 \ln^22}{16})_{lbl}+\frac{839}{144}-\frac{11 \ln2}{24}\big]n_h+(\frac{839}{144}-\frac{11 \ln2}{24})n_l\\&-\frac{3997}{48}+\frac{691 \pi ^2}{192}+\frac{4777 \ln^22}{96}-\frac{1859 \ln2}{48}+i(\frac{203 \pi }{6}-\frac{1735}{48} \pi  \ln2)\bigg\},
\end{aligned}&\\ \notag
&\begin{aligned}
f^{1\,(2)}_{1,1}(0,2) &=\frac{1}{\sqrt{6}}\bigg\{\big[(\frac{5 \ln2}{8}-\frac{5 \ln^22}{8})_{lbl}+\frac{469}{144}+\frac{23 \ln2}{12}\big]n_h+(\frac{469}{144}+\frac{23 \ln2}{12})n_l\\&+\frac{97 \pi ^2}{36}-\frac{457}{16}+\frac{1013 \ln^22}{24}-\frac{649 \ln2}{12}+i( \frac{247 \pi }{24}-\frac{451}{24} \pi  \ln2)\bigg\},
\end{aligned}&\\ \notag
&\begin{aligned}
f^{2\,(2)}_{1,2}(0,2) &=\frac{1}{\sqrt{3}}\bigg\{\big[(-\frac{5}{16}-\frac{15 \ln^22}{8}+\frac{5 \ln2}{4})_{lbl}-\frac{19}{32}-\frac{29 \ln2}{12}-\frac{i \pi }{4}]n_h\\&+(-\frac{137}{96}-\frac{41 \ln2}{12}-\frac{i \pi }{4})n_l+\frac{1705}{48}-\frac{1727 \pi ^2}{288}-\frac{230 \ln^22}{3}+\frac{367 \ln2}{4} \\&+i(\frac{467}{12} \pi  \ln2-\frac{479 \pi }{48})\bigg\},
\end{aligned}&\\ \notag
&\begin{aligned}
f^{2\,(2)}_{1,1}(0,2) &=\frac{1}{\sqrt{6}}\bigg\{[(\frac{5 \ln^22}{8}-\frac{5 \ln2}{8})_{lbl}+\frac{469}{144}-\frac{5 \ln2}{4}\big]n_h+(\frac{469}{144}-\frac{5 \ln2}{4})n_l\\&-\frac{1085}{16}+\frac{139 \pi ^2}{24}+\frac{325 \ln^22}{8}-\frac{373 \ln2}{12}+i(\frac{135 \pi }{8}-\frac{545}{24} \pi  \ln2)\bigg\},
\end{aligned}&\\ \notag
&\begin{aligned}
f^{2\,(2)}_{1,0}(0,2) &=\frac{1}{\sqrt{2}}\bigg\{\big[(\frac{5 \ln2}{24}-\frac{5 \ln^22}{12})_{lbl}+\frac{7}{6}+\frac{25 \ln2}{18}\big]n_h+(\frac{7}{6}+\frac{25 \ln2}{18})n_l\\&+\frac{193}{36}-\frac{299 \pi ^2}{864}+\frac{1049 \ln^22}{144}-\frac{251 \ln2}{9}+i(\frac{395 \pi }{72}-\frac{31}{36} \pi  \ln2)\bigg\},
\end{aligned}&\\ \notag
&\begin{aligned}
f^{2\,(2)}_{0,1}(0,2) &=\frac{1}{\sqrt{6}}\bigg\{\big[(\frac{5 \ln2}{16}-\frac{5 \ln^22}{16})_{lbl}+\frac{527}{144}-\frac{\ln2}{8}\big]n_h+(\frac{527}{144}-\frac{\ln2}{8})n_l\\&-\frac{683}{16}+\frac{193 \pi ^2}{64}+\frac{2969 \ln^22}{96}-\frac{1771 \ln2}{48}+i( 16 \pi -\frac{237}{16} \pi  \ln2)\bigg\},
\end{aligned}&\\ \notag
&\begin{aligned}
f^{2\,(2)}_{0,0}(0,2) &=\frac{1}{\sqrt{2}}\bigg[(\frac{1}{2}-\frac{\pi ^2}{4}-\ln^22+\frac{4 \ln2}{3})C_AT_F+(\frac{11 \ln2}{12}-\frac{11}{12})C_AC_F\\& \notag+( -\frac{1}{2}+\frac{\pi ^2}{24}-\frac{3 \ln^22}{4}+\frac{7 \ln2}{4})C_F^2+\big[-1+\frac{\pi ^2}{12}+3 \ln^22-3 \ln2\\&+(\frac{1}{3}-\frac{\ln2}{3})n_h+(\frac{1}{3}-\frac{\ln2}{3})n_l\big]C_FT_F+(\frac{4}{3} \ln2\,n_h+\frac{4}{3} \ln2\,n_l)T_F^2\bigg],
\end{aligned}&\\
\end{flalign}

\begin{flalign}
&\begin{aligned}
f^{0\,(2)}_{1,0}(0,1) &=\big[\frac{335 \zeta (3)}{64}-\frac{5}{48}+\frac{5 \pi ^2}{32}-\frac{5 \ln^32}{4}+\frac{25 \ln^22}{12}-\frac{35 \ln2}{4}-\frac{5}{16} \pi ^2 \ln2 \\&+i(\frac{5}{12} \pi  \ln^22-\frac{5}{24} \pi  \ln2)\big]_{lbl}n_h\\&+\big[\frac{1279}{216}+\frac{377 \pi ^2}{216}+\frac{181 \ln^22}{144}+\frac{3185 \ln2}{432}+i( -\frac{79 \pi }{36}-\frac{29}{18} \pi  \ln2)\big]n_h\\&+\big[\frac{1279}{216}+\frac{377 \pi ^2}{216}+\frac{181 \ln^22}{144}+\frac{3185 \ln2}{432} +i(-\frac{79 \pi }{36}-\frac{29}{18} \pi  \ln2)\big]n_l-\frac{19921 \zeta (3)}{576}\\&-\frac{27797}{432}-\frac{10363 \pi ^2}{576}+\frac{7889 \ln^32}{216}-\frac{1963 \ln^22}{72}-\frac{26339 \ln2}{216}-\frac{145}{864} \pi ^2 \ln2\\&+i(\frac{505 \pi }{18}-\frac{565 \pi ^3}{864}-\frac{4889}{144} \pi  \ln^22+\frac{4753}{72} \pi  \ln2), \notag
\end{aligned}&\\ \notag
&\begin{aligned}
f^{0\,(2)}_{0,0}(0,1) &=\big[\frac{43}{54}+\frac{\pi ^2}{24}-\frac{\ln^22}{9}+\frac{31 \ln2}{27}+i(\frac{2 \pi }{9}-\frac{2}{9} \pi  \ln2 )\big]n_h\\&+\big[\frac{43}{54}+\frac{\pi ^2}{24}-\frac{\ln^22}{9}+\frac{31 \ln2}{27}+i(\frac{2 \pi }{9}-\frac{2}{9} \pi  \ln2)\big]n_l\\&-\frac{4147 \zeta (3)}{576}-\frac{3395}{324}+\frac{23 \pi ^2}{648}-\frac{89 \ln^32}{54}+\frac{295 \ln^22}{36}-\frac{253 \ln2}{324}-\frac{185}{432}
   \pi ^2 \ln2\\&+i( -\frac{103 \pi }{36}+\frac{53 \pi ^3}{216}+\frac{5}{6} \pi  \ln^22-\frac{89}{18} \pi  \ln2),
\end{aligned}&\\ \notag
&\begin{aligned}
f^{1\,(2)}_{1,0}(0,1) &=\frac{1}{\sqrt{6}}\bigg\{\big[ \frac{1905 \zeta (3)}{128}-\frac{5}{6}+\frac{15 \pi ^2}{32}-\frac{15 \ln^32}{4}+\frac{85 \ln^22}{16}-\frac{145 \ln2}{6}-\frac{15}{16} \pi ^2 \ln2\\&+i( \frac{5}{4} \pi  \ln^22-\frac{5 \pi }{8})\big]_{lbl}n_h\\&+\big[\frac{13}{72}+\frac{17 \pi ^2}{12}+\frac{63 \ln^22}{16}+\frac{455 \ln2}{72}+i(\frac{1}{6} \pi  \ln2-\frac{167 \pi }{72})\big]n_h\\&+\big[\frac{13}{72}+\frac{17 \pi ^2}{12}+\frac{63 \ln^22}{16}+\frac{455 \ln2}{72}+i(\frac{1}{6} \pi  \ln2-\frac{167 \pi }{72})\big]n_l\\&-\frac{1803 \zeta (3)}{32}-\frac{469}{36}-\frac{2843 \pi ^2}{288}+\frac{2231 \ln^32}{72}-\frac{353 \ln^22}{8}-\frac{6817 \ln2}{72}+\frac{13}{18} \pi
   ^2 \ln2\\&+i(-\frac{\pi }{24}+\frac{11 \pi ^3}{36}-\frac{133}{4} \pi  \ln^22+\frac{135}{2} \pi  \ln2)\bigg\},
\end{aligned}&\\ \notag
&\begin{aligned}
f^{1\,(2)}_{0,1}(0,1) &=\frac{1}{\sqrt{6}}\bigg\{\big[\frac{3855 \zeta (3)}{256}+\frac{15}{16}+\frac{15 \pi ^2}{64}+\frac{15 \ln^32}{8}-\frac{175 \ln^22}{32}-\frac{385 \ln2}{16}-\frac{35}{96} \pi ^2
   \ln2 \\&+i(\frac{5}{8} \pi  \ln2-\frac{5}{8} \pi  \ln^22)\big]_{lbl}n_h+\big[ -\frac{2281}{144}-\frac{205 \pi ^2}{72}+\frac{427 \ln^22}{48}-\frac{169 \ln2}{72}\\&+i(\frac{287 \pi }{72}+\frac{29}{4} \pi  \ln2)\big]n_h+\big[ -\frac{2281}{144}-\frac{205 \pi ^2}{72}+\frac{427 \ln^22}{48}-\frac{169 \ln2}{72}\\&+i(\frac{287 \pi }{72}+\frac{29}{4} \pi  \ln2)\big]n_l+\frac{56005 \zeta (3)}{384}+\frac{4324}{27}+\frac{401 \pi ^2}{864}+\frac{3473 \ln^32}{144}\\&-\frac{3625 \ln^22}{8}+\frac{17149 \ln2}{54}+\frac{6883}{288} \pi ^2 \ln2+i(-\frac{961 \pi }{24}+\frac{1369 \pi ^3}{288}\\&+\frac{4777}{48} \pi  \ln^22-\frac{5093}{24} \pi  \ln2)\bigg\},
\end{aligned}&\\ \notag
&\begin{aligned}
f^{1\,(2)}_{1,1}(0,1) &=\frac{1}{\sqrt{6}}\bigg\{\big[-\frac{4155 \zeta (3)}{128}-\frac{25}{12}-\frac{125 \pi ^2}{96}+\frac{15 \ln^32}{4}-\frac{195 \ln^22}{16}+\frac{1625 \ln2}{24}\\&+\frac{85}{48} \pi ^2
   \ln2+i(\frac{5}{4} \pi  \ln2-\frac{5}{4} \pi  \ln^22)\big]_{lbl}n_h+\big[\frac{421}{72}-\frac{11 \pi ^2}{4}-\frac{31 \ln^22}{16}-\frac{65 \ln2}{8}\\&+i(\frac{493 \pi }{72}+\frac{17}{6} \pi  \ln2)\big]n_h+\big[\frac{421}{72}-\frac{11 \pi ^2}{4}-\frac{31 \ln^22}{16}-\frac{65 \ln2}{8}\\&+i(\frac{493 \pi }{72}+\frac{17}{6} \pi  \ln2 )\big]n_l+\frac{14035 \zeta (3)}{192}-\frac{3337}{54}+\frac{25085 \pi ^2}{864}-\frac{4669 \ln^32}{72}\\&+\frac{1213 \ln^22}{24}+\frac{1009 \ln2}{54}+\frac{349}{48} \pi ^2 \ln2\\&+i(-\frac{501 \pi }{8}+\frac{65 \pi ^3}{18}+\frac{1013}{12} \pi  \ln^22-\frac{275}{3} \pi  \ln2)\bigg\},
\end{aligned}&\\ \notag
&\begin{aligned}
f^{2\,(2)}_{1,2}(0,1) &=\frac{1}{\sqrt{3}}\bigg\{\big[ \frac{3345 \zeta (3)}{128}-\frac{55 \pi ^2}{96}+\frac{45 \ln^32}{4}-\frac{575 \ln^22}{16}-\frac{45 \ln2}{2}+\frac{5}{16} \pi ^2 \ln2 \\&+i( -\frac{5 \pi }{8}-\frac{15}{4} \pi  \ln^22+\frac{5}{2} \pi  \ln2)\big]_{lbl}n_h+\big[-\frac{271}{144}-\frac{7 \pi ^2}{36}-\frac{37 \ln^22}{24}+\frac{203 \ln2}{72}\\&+i(\frac{17}{6} \pi  \ln2-\frac{161 \pi }{48})\big]n_h+\big[-\frac{143}{144}+\frac{5 \pi ^2}{36}+\frac{35 \ln^22}{24}+\frac{83 \ln2}{72}\\&+i(\frac{5}{6} \pi  \ln2-\frac{27 \pi }{16})\big]n_l+ \frac{419 \zeta (3)}{12}+\frac{6647}{864}+\frac{157 \pi ^2}{27}+\frac{9037 \ln^32}{72}\\&-\frac{7235 \ln^22}{24}-\frac{14563 \ln2}{432}-\frac{671}{48}
   \pi ^2 \ln2+i(\frac{2669 \pi }{72}-\frac{1165 \pi ^3}{144}\\&-\frac{2987}{24} \pi  \ln^22+\frac{1235}{12} \pi  \ln2)\bigg\},
\end{aligned}&\\ \notag
&\begin{aligned}
f^{2\,(2)}_{1,1}(0,1) &=\frac{1}{\sqrt{6}}\bigg\{\big[ \frac{1275 \zeta (3)}{128}-\frac{5}{4}+\frac{125 \pi ^2}{96}-\frac{15 \ln^32}{4}+\frac{475 \ln^22}{16}-\frac{355 \ln2}{8}-\frac{15}{16} \pi ^2 \ln2\\&+i(\frac{5}{4} \pi  \ln^22-\frac{5}{4} \pi  \ln2)\big]_{lbl}n_h+\big[ -\frac{1133}{72}-\frac{11 \pi ^2}{12}+\frac{413 \ln^22}{48}+\frac{383 \ln2}{72}\\&+i(\frac{31}{6} \pi  \ln2-\frac{83 \pi }{72})\big]n_h+\big[-\frac{1133}{72}-\frac{11 \pi ^2}{12}+\frac{413 \ln^22}{48}+\frac{383 \ln2}{72}\\&+i(\frac{31}{6} \pi  \ln2-\frac{83 \pi }{72})\big]n_l+ -\frac{11625 \zeta (3)}{64}+\frac{9085}{54}+\frac{11683 \pi ^2}{864}-\frac{155 \ln^32}{72}-\frac{877 \ln^22}{24}\\&-\frac{9995 \ln2}{108}+\frac{1039}{48} \pi ^2 \ln2+i(-\frac{211 \pi }{24}+\frac{385 \pi ^3}{36}+\frac{325}{4} \pi  \ln^22-\frac{1133}{6} \pi  \ln2)\bigg\},
\end{aligned}&\\ \notag
&\begin{aligned}
f^{2\,(2)}_{1,0}(0,1) &=\frac{1}{\sqrt{2}}\bigg\{\big[ -\frac{335 \zeta (3)}{32}+\frac{5}{24}-\frac{5 \pi ^2}{16}+\frac{5 \ln^32}{2}-\frac{25 \ln^22}{6}+\frac{35 \ln2}{2}+\frac{5}{8} \pi ^2 \ln2\\&+i(\frac{5}{12} \pi  \ln2-\frac{5}{6} \pi  \ln^22)\big]_{lbl}n_h+\big[ -\frac{271}{108}-\frac{137 \pi ^2}{108}-\frac{181 \ln^22}{72}+\frac{559 \ln2}{216} \\&+i(\frac{41 \pi }{18}+\frac{5}{9} \pi  \ln2)\big]n_h+\big[-\frac{271}{108}-\frac{137 \pi ^2}{108}-\frac{181 \ln^22}{72}+\frac{559 \ln2}{216}\\&+i(\frac{41 \pi }{18}+\frac{5}{9} \pi  \ln2)\big]n_l+\frac{13585 \zeta (3)}{288}+\frac{11405}{216}+\frac{5563 \pi ^2}{288}-\frac{3281 \ln^32}{108}\\&+\frac{5711 \ln^22}{72}-\frac{12311 \ln2}{54}-\frac{1391}{432} \pi ^2 \ln2+i( \frac{95 \pi }{12}-\frac{587 \pi ^3}{432}+\frac{1049}{72} \pi  \ln^22\\&-\frac{691}{36} \pi  \ln2)\bigg\},
\end{aligned}&\\ \notag
&\begin{aligned}
f^{2\,(2)}_{0,1}(0,1) &=\frac{1}{\sqrt{6}}\bigg\{\big[\frac{3855 \zeta (3)}{256}+\frac{15}{16}+\frac{15 \pi ^2}{64}+\frac{15 \ln^32}{8}-\frac{175 \ln^22}{32}-\frac{385 \ln2}{16}-\frac{35}{96} \pi ^2
   \ln2 \\&+i(\frac{5}{8} \pi  \ln2-\frac{5}{8} \pi  \ln^22 )\big]_{lbl}n_h+\big[-\frac{1513}{144}-\frac{35 \pi ^2}{24}+\frac{57 \ln^22}{16}+\frac{7 \ln2}{72}\\&+i(\frac{167 \pi }{72}+\frac{31}{12} \pi  \ln2)\big]n_h+\big[-\frac{1513}{144}-\frac{35 \pi ^2}{24}+\frac{57 \ln^22}{16}+\frac{7 \ln2}{72}\\&+i(\frac{167 \pi }{72}+\frac{31}{12} \pi  \ln2)\big]n_l-\frac{15379 \zeta (3)}{384}+\frac{5965}{54}+\frac{21817 \pi ^2}{864}-\frac{391 \ln^32}{16}\\&-\frac{369 \ln^22}{8}-\frac{50 \ln2}{27}+\frac{49}{96}
   \pi ^2 \ln2+i(-\frac{1159 \pi }{72}+\frac{1417 \pi ^3}{288}+\frac{2969}{48} \pi  \ln^22\\&-\frac{943}{8} \pi  \ln2)\bigg\},
\end{aligned}&\\ \notag
&\begin{aligned}
f^{2\,(2)}_{0,0}(0,1) &=\frac{1}{\sqrt{2}}\bigg\{\big[-\frac{43}{27}-\frac{\pi ^2}{12}+\frac{2 \ln^22}{9}+\frac{46 \ln2}{27}+i(\frac{4}{9} \pi  \ln2-\frac{7 \pi }{9})\big]n_h\\&+\big[-\frac{43}{27}-\frac{\pi ^2}{12}+\frac{2 \ln^22}{9}+\frac{46 \ln2}{27}+i(\frac{4}{9} \pi  \ln2-\frac{7 \pi }{9})\big]n_l\\&+ \frac{4147 \zeta (3)}{288}+\frac{5863}{324}-\frac{134 \pi ^2}{81}+\frac{89 \ln^32}{27}-\frac{295 \ln^22}{18}-\frac{5399 \ln2}{162}+\frac{185}{216}
   \pi ^2 \ln2\\&+i( \frac{119 \pi }{12}-\frac{53 \pi ^3}{108}-\frac{5}{3} \pi  \ln^22+\frac{83}{9} \pi  \ln2)\bigg\},
\end{aligned}&\\
\end{flalign}

\begin{flalign}
&\begin{aligned}
f^{0\,(2)}_{1,0}(0,0) &=\big[-0.987311+i(\frac{335 \pi  \zeta (3)}{64}-\frac{5 \pi }{48}+\frac{5 \pi ^3}{32}-\frac{5}{4} \pi  \ln^32+\frac{25}{12} \pi  \ln^22\\&-\frac{35}{4} \pi  \ln2-\frac{5}{16} \pi ^3 \ln2)\big]_{lbl}n_h+\big[3.95524+i( -\frac{1265 \pi }{54}+\frac{121 \pi ^3}{216}\\&+\frac{1237}{144} \pi  \ln^22-\frac{6091}{432} \pi  \ln2)\big]n_h+\big[-35.3031+i(-\frac{637 \pi }{27}+\frac{115 \pi ^3}{216}\\&+\frac{1237}{144} \pi  \ln^22-\frac{6091}{432} \pi  \ln2)\big]n_l+(-3.17554 +\frac{17i \pi }{9})n_hn_l\\&+\big[-1.58777+\frac{17i \pi }{18}\big]n_h^2+\big[-1.58777+\frac{17 i\pi }{18}\big]n_l^2-148.113\\&+i(-\frac{7517 \pi  \zeta (3)}{288}+\frac{44971 \pi }{432}-\frac{1021 \pi ^3}{144}+\frac{7889}{216} \pi  \ln^32-\frac{10675}{72} \pi  \ln^22\\&+\frac{12547}{54} \pi  \ln2+\frac{3679}{864} \pi ^3 \ln2), \notag
\end{aligned}&\\ \notag
&\begin{aligned}
f^{0\,(2)}_{0,0}(0,0) &=\big[(-0.183540)_{lbl}-0.718592+i(-\frac{871 \pi }{216}+\frac{13 \pi ^3}{216}+\frac{13}{18} \pi  \ln^22-\frac{13}{27} \pi  \ln2)\big]n_h\\&+\big[-4.05636+i(-\frac{907 \pi }{216}+\frac{7 \pi ^3}{216}+\frac{13}{18} \pi  \ln^22-\frac{13}{27} \pi  \ln2)\big]n_l+(-0.208805+\frac{8 i\pi }{27})n_h n_l\\&+(-0.104403+\frac{4 i\pi }{27})n_h^2+( -0.104403+\frac{4 i\pi }{27})n_l^2-20.0331+i(\frac{1145 \pi  \zeta (3)}{576}\\&+\frac{30763 \pi }{1296}-\frac{1781 \pi ^3}{1296}-\frac{89}{54} \pi  \ln^32-\frac{50}{9} \pi  \ln^22+\frac{8459}{324} \pi
    \ln2-\frac{23}{432} \pi ^3 \ln2),
\end{aligned}&\\ \notag
&\begin{aligned}
f^{1\,(2)}_{1,0}(0,0) &=\frac{1}{\sqrt{6}}\bigg\{\big[0.447507+i(\frac{1905 \pi  \zeta (3)}{128}-\frac{5 \pi }{6}+\frac{15 \pi ^3}{32}-\frac{15}{4} \pi  \ln^32+\frac{85}{16} \pi  \ln^22\\&-\frac{145}{6} \pi  \ln2-\frac{15}{16} \pi ^3 \ln2)\big]_{lbl}n_h+\big[-13.1562+i(\frac{241 \pi }{72}-\frac{\pi ^3}{36}+\frac{55}{16} \pi  \ln^22\\&-\frac{277}{72} \pi  \ln2)\big]n_h+\big[-32.8520+i( \frac{241 \pi }{72}-\frac{\pi ^3}{36}+\frac{55}{16} \pi  \ln^22-\frac{277}{72} \pi  \ln2)\big]n_l\\&+419.991+i(-\frac{969 \pi  \zeta (3)}{16}-\frac{4715 \pi }{72}+\frac{511 \pi ^3}{144}+\frac{2231}{72} \pi  \ln^32-\frac{287}{8} \pi  \ln^22\\&+\frac{5261}{72} \pi 
   \ln2+\frac{13}{2} \pi ^3 \ln2)\bigg\},
\end{aligned}&\\ \notag
&\begin{aligned}
f^{1\,(2)}_{0,1}(0,0) &=\frac{1}{\sqrt{6}}\bigg\{\big[-0.746677+i(\frac{3855 \pi  \zeta (3)}{256}+\frac{15 \pi }{16}+\frac{15 \pi ^3}{64}+\frac{15}{8} \pi  \ln^32-\frac{175}{32} \pi  \ln^22\\&-\frac{385}{16} \pi  \ln2-\frac{35}{96} \pi ^3 \ln2)\big]_{lbl}n_h+\big[-43.1373+i(\frac{3487 \pi }{144}-\frac{17 \pi ^3}{8}-\frac{387}{16} \pi  \ln^22\\&+\frac{4327}{72} \pi  \ln2)\big]n_h+\big[3.39730+i( \frac{3703 \pi }{144}-\frac{55 \pi ^3}{24}-\frac{387}{16} \pi  \ln^22+\frac{4327}{72} \pi  \ln2)\big]n_l\\&+(3.61677-\frac{26i \pi }{9})n_hn_l+(1.80839-\frac{13i \pi }{9})n_h^2+(1.80839-\frac{13i \pi }{9})n_l^2+844.029\\&+i(\frac{52657 \pi  \zeta (3)}{384}-\frac{24613 \pi }{864}+\frac{9269 \pi ^3}{864}+\frac{3473}{144} \pi  \ln^32+\frac{371}{4} \pi  \ln^22-\frac{38489}{54}
   \pi  \ln2\\&+\frac{197}{96} \pi ^3 \ln2)\bigg\},
\end{aligned}&\\ \notag
&\begin{aligned}
f^{1\,(2)}_{1,1}(0,0) &=\frac{1}{\sqrt{6}}\bigg\{\big[-0.800531+i(-\frac{4155 \pi  \zeta (3)}{128}-\frac{25 \pi }{12}-\frac{125 \pi ^3}{96}+\frac{15}{4} \pi  \ln^32-\frac{195}{16} \pi  \ln^22\\&+\frac{1625}{24} \pi  \ln2+\frac{85}{48} \pi ^3 \ln2)\big]_{lbl}n_h+\big[-72.1677+i(\frac{3473 \pi }{72}-\frac{11 \pi ^3}{12}-\frac{709}{48} \pi  \ln^22\\&-\frac{121}{72} \pi  \ln2)\big]n_h+\big[-39.6679+i(\frac{3473 \pi
   }{72}-\frac{11 \pi ^3}{12}-\frac{709}{48} \pi  \log
   ^2(2)-\frac{121}{72} \pi  \log (2))\big]n_l\\&+(3.61677-\frac{26 i\pi }{9})n_hn_l+(1.80839-\frac{13 i\pi }{9})n_h^2\\&+(1.80839-\frac{13i \pi }{9})n_l^2+1416.64+i(\frac{14785 \pi  \zeta (3)}{192}-\frac{62461 \pi }{216}+\frac{2341 \pi ^3}{432}\\&-\frac{4669}{72} \pi  \ln^32+\frac{6295}{24} \pi  \ln^22-\frac{4733}{54}
   \pi  \ln2-\frac{757}{144} \pi ^3 \ln2)\bigg\},
\end{aligned}&\\ \notag
&\begin{aligned}
f^{2\,(2)}_{1,2}(0,0) &=\frac{1}{\sqrt{3}}\bigg\{\big[1.84256+i(\frac{3345 \pi  \zeta (3)}{128}-\frac{55 \pi ^3}{96}+\frac{45}{4} \pi  \ln^32-\frac{575}{16} \pi  \ln^22\\&-\frac{45}{2} \pi  \ln2+\frac{5}{16} \pi ^3 \ln2)\big]_{lbl}n_h+\big[-10.5174+i(\frac{1933 \pi }{144}+\frac{\pi ^3}{36}-\frac{433}{24} \pi  \ln^22\\&+\frac{931}{72} \pi  \ln2)\big]n_h+\big[-2.20789+i(\frac{2837 \pi }{144}-\frac{\pi ^3}{18}-\frac{265}{24} \pi  \ln^22+\frac{67}{72} \pi  \ln2)\big]n_l\\&+(2.36394-\frac{10 i\pi }{9})n_hn_l+(1.18197-\frac{5i \pi }{9})n_h^2+(1.18197-\frac{5 i\pi }{9})n_l^2\\&+ 685.230+i(\frac{26413 \pi  \zeta (3)}{384}-\frac{339427 \pi }{1728}+\frac{179393 \pi ^3}{60480}+\frac{839}{9} \pi  \ln^32\\&-\frac{493}{6} \pi  \ln^22-\frac{1462}{27} \pi  \ln2-\frac{563}{144} \pi ^3 \ln2)\bigg\},
\end{aligned}&\\ \notag
&\begin{aligned}
f^{2\,(2)}_{1,1}(0,0) &=\frac{1}{\sqrt{6}}\bigg\{\big[0.757694+i( \frac{1275 \pi  \zeta (3)}{128}-\frac{5 \pi }{4}+\frac{125 \pi ^3}{96}-\frac{15}{4} \pi  \ln^32+\frac{475}{16} \pi  \ln^22\\&-\frac{355}{8} \pi  \ln2-\frac{15}{16} \pi ^3 \ln2)\big]_{lbl}n_h+\big[-37.5140+i(\frac{629 \pi }{24}-\frac{55 \pi ^3}{36}-\frac{475}{48} \pi  \ln^22\\&+\frac{917}{24} \pi  \ln2)\big]n_h+\big[-22.4630+i(\frac{629 \pi }{24}-\frac{55 \pi ^3}{36}-\frac{475}{48} \pi  \ln^22+\frac{917}{24} \pi  \ln2)\big]n_l\\&+(3.61677-\frac{26i \pi }{9})n_hn_l+(1.80839 -\frac{13i \pi }{9})n_h^2+(1.80839 -\frac{13i \pi }{9})n_l^2\\&+996.866+i(-\frac{9691 \pi  \zeta (3)}{64}-\frac{13691 \pi }{216}+\frac{514837 \pi ^3}{15120}-\frac{155}{72} \pi  \ln^32\\&+\frac{6295}{24} \pi  \ln^22-\frac{17666}{27} \pi  \ln2+\frac{937}{144} \pi ^3 \ln2)\bigg\},
\end{aligned}&\\ \notag
&\begin{aligned}
f^{2\,(2)}_{1,0}(0,0) &=\frac{1}{\sqrt{2}}\bigg\{\big[0.648250+i(-\frac{335 \pi  \zeta (3)}{32}+\frac{5 \pi }{24}-\frac{5 \pi ^3}{16}+\frac{5}{2} \pi  \ln^32-\frac{25}{6} \pi  \ln^22\\&+\frac{35}{2} \pi  \ln2+\frac{5}{8} \pi ^3 \ln2)\big]_{lbl}n_h+\big[-23.9086+i(\frac{608 \pi }{27}-\frac{25 \pi ^3}{108}\\&-\frac{469}{72} \pi  \ln^22-\frac{101}{216} \pi  \ln2)\big]n_h+\big[-9.87109+i(\frac{614 \pi }{27}-\frac{19 \pi ^3}{108}\\&-\frac{469}{72} \pi  \ln^22-\frac{245}{216} \pi  \ln2)\big]n_l+(1.62320-\frac{14 i\pi }{9})n_hn_l+(0.811601-\frac{7 i\pi }{9})n_h^2\\&+(0.811601-\frac{7 i\pi }{9})n_l^2+486.981+i(\frac{2269 \pi  \zeta (3)}{72}-\frac{92927 \pi }{864}+\frac{13043 \pi ^3}{1680}\\&-\frac{3281}{108} \pi  \ln^32+\frac{1351}{9} \pi  \ln^22-\frac{18385}{108} \pi  \ln2-\frac{2143}{432} \pi ^3 \ln2)\bigg\},
\end{aligned}&\\ \notag
&\begin{aligned}
f^{2\,(2)}_{0,1}(0,0) &=\frac{1}{\sqrt{6}}\bigg\{\big[0.0744372+i(\frac{3855 \pi  \zeta (3)}{256}+\frac{15 \pi }{16}+\frac{15 \pi ^3}{64}+\frac{15}{8} \pi  \ln^32-\frac{175}{32} \pi  \ln^22\\&-\frac{385}{16} \pi  \ln2-\frac{35}{96} \pi ^3 \ln2)\big]_{lbl}n_h+\big[-32.0880+i(\frac{251 \pi }{16}-\frac{61 \pi ^3}{72}-\frac{521}{48} \pi  \ln^22\\&+\frac{1487}{72} \pi  \ln2)\big]n_h+\big[-13.1035+i( \frac{2987 \pi }{144}-\frac{73 \pi ^3}{72}-\frac{521}{48} \pi  \ln^22+\frac{1295}{72} \pi  \ln2)\big]n_l\\&+(1.25283-\frac{16 i\pi }{9})n_hn_l+(0.626416-\frac{8 i\pi }{9})n_h^2+(0.626416-\frac{8 i\pi }{9})n_l^2\\&+507.654+i(\frac{18461 \pi  \zeta (3)}{384}-\frac{21637 \pi }{96}+\frac{172489 \pi ^3}{6048}-\frac{391}{16} \pi  \ln^32+\frac{2333}{12} \pi  \ln^22\\&-\frac{31151}{108} \pi  \ln2-\frac{683}{96} \pi ^3 \ln2)\bigg\},
\end{aligned}&\\ \notag
&\begin{aligned}
f^{2\,(2)}_{0,0}(0,0) &=\frac{1}{\sqrt{2}}\bigg\{(0.367080)_{lbl}n_h+\big[-7.24328+i(\frac{799 \pi }{108}-\frac{13 \pi ^3}{108}-\frac{13}{9} \pi  \ln^22+\frac{8}{27} \pi  \ln2)\big]n_h\\&+\big[-8.05584+i(\frac{823 \pi }{108}-\frac{7 \pi ^3}{108}-\frac{13}{9} \pi  \ln^22-\frac{10}{27} \pi  \ln2)\big]n_l+(0.417610-\frac{16i \pi }{27})n_hn_l\\&+(0.208805-\frac{8i \pi }{27})n_h^2+(0.208805-\frac{8i \pi }{27})n_l^2+203.973+i(\frac{845 \pi  \zeta (3)}{144}-\frac{81025 \pi }{2592}\\&+\frac{18623 \pi ^3}{45360}+\frac{89}{27} \pi  \ln^32+\frac{121}{9} \pi  \ln^22-\frac{6521}{648}
   \pi  \ln2+\frac{23}{216} \pi ^3 \ln2)\bigg\}.
\end{aligned}&\\ 
\end{flalign}

\bibliography{biblio.bib}

@article{Belle:2002tfa,
    author = "Abe, Kazuo and others",
    collaboration = "Belle",
    title = "{Observation of double c anti-c production in e+ e- annihilation at s**(1/2) approximately 10.6-GeV}",
    eprint = "hep-ex/0205104",
    archivePrefix = "arXiv",
    reportNumber = "BELLE-PREPRINT-2002-13, KEK-PREPRINT-2002-27",
    doi = "10.1103/PhysRevLett.89.142001",
    journal = "Phys. Rev. Lett.",
    volume = "89",
    pages = "142001",
    year = "2002"
}

@article{BaBar:2005nic,
    author = "Aubert, Bernard and others",
    editor = "Brenner, R. and de los Heros, C. P. and Rathsman, J.",
    collaboration = "BaBar",
    title = "{Measurement of double charmonium production in $e^+e^-$ annihilations at $\sqrt{s}=10.6$ GeV}",
    eprint = "hep-ex/0506062",
    archivePrefix = "arXiv",
    reportNumber = "BABAR-PUB-05-21, SLAC-PUB-11287, BABAR-PUB-05-021",
    doi = "10.1103/PhysRevD.72.031101",
    journal = "Phys. Rev. D",
    volume = "72",
    pages = "031101",
    year = "2005"
}

@article{Braaten:2002fi,
    author = "Braaten, Eric and Lee, Jungil",
    title = "{Exclusive Double Charmonium Production from $e^+ e^-$ Annihilation into a Virtual Photon}",
    eprint = "hep-ph/0211085",
    archivePrefix = "arXiv",
    reportNumber = "FERMILAB-PUB-02-274-T, ANL-HEP-PR-02-094",
    doi = "10.1103/PhysRevD.72.099901",
    journal = "Phys. Rev. D",
    volume = "67",
    pages = "054007",
    year = "2003",
    note = "[Erratum: Phys.Rev.D 72, 099901 (2005)]"
}

@article{Hagiwara:2003cw,
    author = "Hagiwara, Kaoru and Kou, Emi and Qiao, Cong-Feng",
    title = "{Exclusive $J/\psi$ productions at $e^{+} e^{-}$ colliders}",
    eprint = "hep-ph/0305102",
    archivePrefix = "arXiv",
    reportNumber = "IPPP-02-71, DCPT-02-141, KEK-TH-871",
    doi = "10.1016/j.physletb.2003.07.006",
    journal = "Phys. Lett. B",
    volume = "570",
    pages = "39--45",
    year = "2003"
}

@article{Liu:2002wq,
    author = "Liu, Kui-Yong and He, Zhi-Guo and Chao, Kuang-Ta",
    title = "{Problems of double charm production in e+ e- annihilation at s**(1/2) = 10.6-GeV}",
    eprint = "hep-ph/0211181",
    archivePrefix = "arXiv",
    doi = "10.1016/S0370-2693(03)00176-X",
    journal = "Phys. Lett. B",
    volume = "557",
    pages = "45--54",
    year = "2003"
}

@article{Zhang:2005cha,
    author = "Zhang, Yu-Jie and Gao, Ying-jia and Chao, Kuang-Ta",
    title = "{Next-to-leading order QCD correction to e+ e- ---{\ensuremath{>}} J / psi + eta(c) at s**(1/2) = 10.6-GeV}",
    eprint = "hep-ph/0506076",
    archivePrefix = "arXiv",
    doi = "10.1103/PhysRevLett.96.092001",
    journal = "Phys. Rev. Lett.",
    volume = "96",
    pages = "092001",
    year = "2006"
}

@article{Gong:2007db,
    author = "Gong, Bin and Wang, Jian-Xiong",
    title = "{QCD corrections to $J/\psi$ plus $\eta_c$ production in $e^{+} e^{-}$ annihilation at $S^{(1/2)}$ = 10.6-GeV}",
    eprint = "0712.4220",
    archivePrefix = "arXiv",
    primaryClass = "hep-ph",
    doi = "10.1103/PhysRevD.77.054028",
    journal = "Phys. Rev. D",
    volume = "77",
    pages = "054028",
    year = "2008"
}

@article{He:2007te,
    author = "He, Zhi-Guo and Fan, Ying and Chao, Kuang-Ta",
    title = "{Relativistic corrections to J/psi exclusive and inclusive double charm production at B factories}",
    eprint = "hep-ph/0702239",
    archivePrefix = "arXiv",
    doi = "10.1103/PhysRevD.75.074011",
    journal = "Phys. Rev. D",
    volume = "75",
    pages = "074011",
    year = "2007"
}

@article{Dong:2012xx,
    author = "Dong, Hai-Rong and Feng, Feng and Jia, Yu",
    title = "{$O(\alpha_s v^2)$ correction to $e^+e^-\to J/\psi+\eta_c$ at $B$ factories}",
    eprint = "1204.4128",
    archivePrefix = "arXiv",
    primaryClass = "hep-ph",
    doi = "10.1103/PhysRevD.85.114018",
    journal = "Phys. Rev. D",
    volume = "85",
    pages = "114018",
    year = "2012"
}

@article{Li:2013otv,
    author = "Li, Xi-Huai and Wang, Jian-Xiong",
    title = "{$O(\alpha_{s}\upsilon^{2})$ correction to $J/\psi$ plus $\eta_c$ production in $e^{+}e^{-}$ annihilation at $\sqrt{s} =$ 10.6 GeV}",
    eprint = "1301.0376",
    archivePrefix = "arXiv",
    primaryClass = "hep-ph",
    doi = "10.1088/1674-1137/38/4/043101",
    journal = "Chin. Phys. C",
    volume = "38",
    pages = "043101",
    year = "2014"
}

@article{Bodwin:2007ga,
    author = "Bodwin, Geoffrey T. and Lee, Jungil and Yu, Chaehyun",
    title = "{Resummation of Relativistic Corrections to e+ e- ---{\ensuremath{>}} J/psi + eta(c)}",
    eprint = "0710.0995",
    archivePrefix = "arXiv",
    primaryClass = "hep-ph",
    reportNumber = "ANL-HEP-PR-07-79",
    doi = "10.1103/PhysRevD.77.094018",
    journal = "Phys. Rev. D",
    volume = "77",
    pages = "094018",
    year = "2008"
}

@article{Belle:2004abn,
    author = "Abe, Kazuo and others",
    collaboration = "Belle",
    title = "{Study of double charmonium production in e+ e- annihilation at s**(1/2) {\textasciitilde} 10.6-GeV}",
    eprint = "hep-ex/0407009",
    archivePrefix = "arXiv",
    doi = "10.1103/PhysRevD.70.071102",
    journal = "Phys. Rev. D",
    volume = "70",
    pages = "071102",
    year = "2004"
}

@article{Zhang:2008gp,
    author = "Zhang, Yu-Jie and Ma, Yan-Qing and Chao, Kuang-Ta",
    title = "{Factorization and NLO QCD correction in $e^+e^- \to J/\psi(\psi(2S))+\chi_{c0}$ at B Factories}",
    eprint = "0802.3655",
    archivePrefix = "arXiv",
    primaryClass = "hep-ph",
    doi = "10.1103/PhysRevD.78.054006",
    journal = "Phys. Rev. D",
    volume = "78",
    pages = "054006",
    year = "2008"
}

@article{Wang:2011qg,
    author = "Wang, Kai and Ma, Yan-Qing and Chao, Kuang-Ta",
    title = "{QCD corrections to $e^+e^- \to J/\psi(\psi(2S))+\chi_{cj}(J=0,1,2)$ at B Factories}",
    eprint = "1107.2646",
    archivePrefix = "arXiv",
    primaryClass = "hep-ph",
    doi = "10.1103/PhysRevD.84.034022",
    journal = "Phys. Rev. D",
    volume = "84",
    pages = "034022",
    year = "2011"
}

@article{Dong:2011fb,
    author = "Dong, Hai-Rong and Feng, Feng and Jia, Yu",
    title = "{$O(\alpha_s)$ corrections to $J/\psi+\chi_{cJ}$ production at $B$ factories}",
    eprint = "1107.4351",
    archivePrefix = "arXiv",
    primaryClass = "hep-ph",
    doi = "10.1007/JHEP10(2011)141",
    journal = "JHEP",
    volume = "10",
    pages = "141",
    year = "2011",
    note = "[Erratum: JHEP 02, 089 (2013)]"
}

@article{Wang:2013vn,
    author = "Wang, Sheng-Quan and Wu, Xing-Gang and Zheng, Xu-Chang and Shen, Jian-Ming and Zhang, Qiong-Lian",
    title = "{$J/\psi +\chi_{cJ}$ Production at the $B$ Factories under the Principle of Maximum Conformality}",
    eprint = "1301.2992",
    archivePrefix = "arXiv",
    primaryClass = "hep-ph",
    doi = "10.1016/j.nuclphysb.2013.09.003",
    journal = "Nucl. Phys. B",
    volume = "876",
    pages = "731--746",
    year = "2013"
}

@article{Jiang:2018wmv,
    author = "Jiang, YingZhao and Sun, Zhan",
    title = "{Further studies on the exclusive productions of $J/\psi +\chi _{cJ}$ ( $J=0,1,2$ ) via $e^+e^-$ annihilation at the $B$ factories}",
    eprint = "1809.09071",
    archivePrefix = "arXiv",
    primaryClass = "hep-ph",
    doi = "10.1140/epjc/s10052-018-6392-x",
    journal = "Eur. Phys. J. C",
    volume = "78",
    number = "11",
    pages = "892",
    year = "2018"
}

@article{Sun:2021tma,
    author = "Sun, Zhan",
    title = "{Next-to-leading-order study of $J/\psi$ angular distributions in $e^{+}e^{-} \to J/\psi+\eta_c,\chi_{cJ}$ at $\sqrt{s} \approx 10.6$GeV}",
    eprint = "2107.02047",
    archivePrefix = "arXiv",
    primaryClass = "hep-ph",
    doi = "10.1007/JHEP09(2021)073",
    journal = "JHEP",
    volume = "09",
    pages = "073",
    year = "2021"
}

@article{Chen:2017pyi,
    author = "Chen, Long-Bin and Liang, Yi and Qiao, Cong-Feng",
    title = "{NNLO QCD corrections to $\gamma + \eta_c(\eta_b)$ exclusive production in electron-positron collision}",
    eprint = "1710.07865",
    archivePrefix = "arXiv",
    primaryClass = "hep-ph",
    doi = "10.1007/JHEP01(2018)091",
    journal = "JHEP",
    volume = "01",
    pages = "091",
    year = "2018"
}

@article{Sang:2020fql,
    author = "Sang, Wen-Long and Feng, Feng and Jia, Yu",
    title = "{Next-to-next-to-leading-order radiative corrections to$e^+e^-\to\chi_{cJ}+\gamma$ at B factory}",
    eprint = "2008.04898",
    archivePrefix = "arXiv",
    primaryClass = "hep-ph",
    doi = "10.1007/JHEP10(2020)098",
    journal = "JHEP",
    volume = "10",
    pages = "098",
    year = "2020"
}

@article{Li:2025pbt,
    author = "Li, Cong and Sang, Wen-Long and Zhang, Hong-Fei",
    title = "{The next-to-next-to-leading-order QCD corrections to $e^{+}e^{-}\to \eta_c/\chi_{cJ} + \gamma$ at B factories}",
    eprint = "2512.04758",
    archivePrefix = "arXiv",
    primaryClass = "hep-ph",
    doi = "10.1007/JHEP05(2026)163",
    journal = "JHEP",
    volume = "05",
    pages = "163",
    year = "2026"
}

@article{Feng:2019zmt,
    author = "Feng, Feng and Jia, Yu and Mo, Zhewen and Sang, Wen-Long and Zhang, Jia-Yue",
    title = "{Next-to-next-to-leading-order QCD corrections to $e^+e^-\to J/\psi\eta_c$ at B factories}",
    eprint = "1901.08447",
    archivePrefix = "arXiv",
    primaryClass = "hep-ph",
    doi = "10.1016/j.physletb.2024.138506",
    journal = "Phys. Lett. B",
    volume = "850",
    pages = "138506",
    year = "2024"
}

@article{Sang:2022kub,
    author = "Sang, Wen-Long and Feng, Feng and Jia, Yu and Mo, Zhewen and Zhang, Jia-Yue",
    title = "{$\mathcal{O}(\alpha^2_s)$ corrections to $J/\psi+\chi_{c0,1,2}$ production at $B$ factories}",
    eprint = "2202.11615",
    archivePrefix = "arXiv",
    primaryClass = "hep-ph",
    doi = "10.1016/j.physletb.2023.138057",
    journal = "Phys. Lett. B",
    volume = "843",
    pages = "138057",
    year = "2023"
}

@article{Huang:2022dfw,
    author = "Huang, Xu-Dong and Gong, Bin and Wang, Jian-Xiong",
    title = "{Next-to-next-to-leading-order QCD corrections to J/{\ensuremath{\psi}} plus {\ensuremath{\eta}}$_{c}$ production at the B factories}",
    eprint = "2212.03631",
    archivePrefix = "arXiv",
    primaryClass = "hep-ph",
    doi = "10.1007/JHEP02(2023)049",
    journal = "JHEP",
    volume = "02",
    pages = "049",
    year = "2023"
}

@article{Li:2025mng,
    author = "Li, Cong and Huang, Xu-Dong and Sang, Wen-Long",
    title = "{Two loop QCD corrections to e+e{\ensuremath{-}}{\textrightarrow}J/{\ensuremath{\psi}}+{\ensuremath{\eta}}c in asymptotic expansion}",
    eprint = "2506.16317",
    archivePrefix = "arXiv",
    primaryClass = "hep-ph",
    doi = "10.1016/j.physletb.2026.140150",
    journal = "Phys. Lett. B",
    volume = "873",
    pages = "140150",
    year = "2026"
}

@article{Chen:2025qgy,
    author = "Chen, Xiang and Guan, Xin and He, Chuan-Qi and Ma, Yan-Qing and Wang, Jian and Zhang, Da-Jiang",
    title = "{Analytical two-loop amplitudes of e+e-{\textrightarrow}J/{\ensuremath{\psi}}+{\ensuremath{\eta}}c at B factories}",
    eprint = "2508.20777",
    archivePrefix = "arXiv",
    primaryClass = "hep-ph",
    doi = "10.1103/93kb-dqz6",
    journal = "Phys. Rev. D",
    volume = "113",
    number = "7",
    pages = "074023",
    year = "2026"
}

@article{Sang:2023liy,
    author = "Sang, Wen-Long and Feng, Feng and Jia, Yu and Mo, Zhewen and Pan, Jichen and Zhang, Jia-Yue",
    title = "{Optimized O({\ensuremath{\alpha}}s2) Correction to Exclusive Double-J/{\ensuremath{\psi}} Production at B Factories}",
    eprint = "2306.11538",
    archivePrefix = "arXiv",
    primaryClass = "hep-ph",
    doi = "10.1103/PhysRevLett.131.161904",
    journal = "Phys. Rev. Lett.",
    volume = "131",
    number = "16",
    pages = "161904",
    year = "2023"
}

@article{Huang:2023pmn,
    author = "Huang, Xu-Dong and Gong, Bin and Niu, Rui-Chang and Yu, Huai-Min and Wang, Jian-Xiong",
    title = "{Next-to-next-to-leading-order QCD corrections to double J/{\ensuremath{\psi}} production at the B factories}",
    eprint = "2311.04751",
    archivePrefix = "arXiv",
    primaryClass = "hep-ph",
    doi = "10.1007/JHEP02(2024)055",
    journal = "JHEP",
    volume = "02",
    pages = "055",
    year = "2024"
}

@article{Bodwin:1994jh,
    author = "Bodwin, Geoffrey T. and Braaten, Eric and Lepage, G. Peter",
    title = "{Rigorous QCD analysis of inclusive annihilation and production of heavy quarkonium}",
    eprint = "hep-ph/9407339",
    archivePrefix = "arXiv",
    reportNumber = "ANL-HEP-PR-94-24, FERMILAB-PUB-94-073-T, NUHEP-TH-94-5",
    doi = "10.1103/PhysRevD.55.5853",
    journal = "Phys. Rev. D",
    volume = "51",
    pages = "1125--1171",
    year = "1995",
    note = "[Erratum: Phys.Rev.D 55, 5853 (1997)]"
}

@article{Petrelli:1997ge,
    author = "Petrelli, Andrea and Cacciari, Matteo and Greco, Mario and Maltoni, Fabio and Mangano, Michelangelo L.",
    title = "{NLO production and decay of quarkonium}",
    eprint = "hep-ph/9707223",
    archivePrefix = "arXiv",
    reportNumber = "CERN-TH-97-142, DESY-97-090",
    doi = "10.1016/S0550-3213(97)00801-8",
    journal = "Nucl. Phys. B",
    volume = "514",
    pages = "245--309",
    year = "1998"
}

@article{Bodwin:2002cfe,
    author = "Bodwin, Geoffrey T. and Petrelli, Andrea",
    title = "{Order-$v^4$ corrections to $S$-wave quarkonium decay}",
    eprint = "hep-ph/0205210",
    archivePrefix = "arXiv",
    reportNumber = "ANL-HEP-PR-12-105, ANL-HEP-PR-02-031",
    doi = "10.1103/PhysRevD.66.094011",
    journal = "Phys. Rev. D",
    volume = "66",
    pages = "094011",
    year = "2002",
    note = "[Erratum: Phys.Rev.D 87, 039902 (2013)]"
}

@article{Zhang:2022nuf,
    author = "Zhang, Yu-Dong and Sang, Wen-Long and Zhang, Hong-Fei",
    title = "{Higher-Order QCD Corrections to $\Upsilon$ Decay into Double Charmonia}",
    eprint = "2205.06124",
    archivePrefix = "arXiv",
    primaryClass = "hep-ph",
    doi = "10.1103/PhysRevLett.129.112002",
    journal = "Phys. Rev. Lett.",
    volume = "129",
    number = "11",
    pages = "112002",
    year = "2022"
}

@article{Zhang:2021ted,
    author = "Zhang, Yu-Dong and Feng, Feng and Sang, Wen-Long and Zhang, Hong-Fei",
    title = "{Next-to-leading-order QCD corrections to a vector bottomonium radiative decay into a charmonium}",
    eprint = "2109.15223",
    archivePrefix = "arXiv",
    primaryClass = "hep-ph",
    doi = "10.1007/JHEP12(2021)189",
    journal = "JHEP",
    volume = "12",
    pages = "189",
    year = "2021"
}

@article{Czarnecki:1997vz,
    author = "Czarnecki, Andrzej and Melnikov, Kirill",
    title = "{Two loop QCD corrections to the heavy quark pair production cross-section in e+ e- annihilation near the threshold}",
    eprint = "hep-ph/9712222",
    archivePrefix = "arXiv",
    reportNumber = "TTP-97-54",
    doi = "10.1103/PhysRevLett.80.2531",
    journal = "Phys. Rev. Lett.",
    volume = "80",
    pages = "2531--2534",
    year = "1998"
}

@article{Beneke:1997jm,
    author = "Beneke, M. and Signer, A. and Smirnov, Vladimir A.",
    title = "{Two loop correction to the leptonic decay of quarkonium}",
    eprint = "hep-ph/9712302",
    archivePrefix = "arXiv",
    reportNumber = "CERN-TH-97-353",
    doi = "10.1103/PhysRevLett.80.2535",
    journal = "Phys. Rev. Lett.",
    volume = "80",
    pages = "2535--2538",
    year = "1998"
}

@article{Hoang:2006ty,
    author = "Hoang, Andre H. and Ruiz-Femenia, Pedro",
    title = "{Heavy pair production currents with general quantum numbers in dimensionally regularized NRQCD}",
    eprint = "hep-ph/0609151",
    archivePrefix = "arXiv",
    reportNumber = "MPP-2006-48",
    doi = "10.1103/PhysRevD.74.114016",
    journal = "Phys. Rev. D",
    volume = "74",
    pages = "114016",
    year = "2006"
}

@article{Sang:2015uxg,
    author = "Sang, Wen-Long and Feng, Feng and Jia, Yu and Liang, Shuang-Ran",
    title = "{Next-to-next-to-leading-order QCD corrections to $\chi_{c0,2}\rightarrow \gamma\gamma$}",
    eprint = "1511.06288",
    archivePrefix = "arXiv",
    primaryClass = "hep-ph",
    doi = "10.1103/PhysRevD.94.111501",
    journal = "Phys. Rev. D",
    volume = "94",
    number = "11",
    pages = "111501",
    year = "2016"
}

@article{Hahn:2000kx,
    author = "Hahn, Thomas",
    title = "{Generating Feynman diagrams and amplitudes with FeynArts 3}",
    eprint = "hep-ph/0012260",
    archivePrefix = "arXiv",
    reportNumber = "KA-TP-23-2000",
    doi = "10.1016/S0010-4655(01)00290-9",
    journal = "Comput. Phys. Commun.",
    volume = "140",
    pages = "418--431",
    year = "2001"
}

@article{Mertig:1990an,
    author = "Mertig, R. and Bohm, M. and Denner, Ansgar",
    title = "{FEYN CALC: Computer algebraic calculation of Feynman amplitudes}",
    reportNumber = "PRINT-90-0639 (WURZBURG)",
    doi = "10.1016/0010-4655(91)90130-D",
    journal = "Comput. Phys. Commun.",
    volume = "64",
    pages = "345--359",
    year = "1991"
}

@article{Feng:2012tk,
    author = "Feng, Feng and Mertig, Rolf",
    title = "{FormLink/FeynCalcFormLink : Embedding FORM in Mathematica and FeynCalc}",
    eprint = "1212.3522",
    archivePrefix = "arXiv",
    primaryClass = "hep-ph",
    month = "12",
    year = "2012"
}

@misc{calcloop,
    author= {} ,
    howpublished = {The CalcLoop package:   \url{https://gitlab.com/multiloop-pku/calcloop}}
}

@article{Liu:2017jxz,
    author = "Liu, Xiao and Ma, Yan-Qing and Wang, Chen-Yu",
    title = "{A Systematic and Efficient Method to Compute Multi-loop Master Integrals}",
    eprint = "1711.09572",
    archivePrefix = "arXiv",
    primaryClass = "hep-ph",
    doi = "10.1016/j.physletb.2018.02.026",
    journal = "Phys. Lett. B",
    volume = "779",
    pages = "353--357",
    year = "2018"
}

@article{Liu:2022chg,
    author = "Liu, Xiao and Ma, Yan-Qing",
    title = "{AMFlow: A Mathematica package for Feynman integrals computation via auxiliary mass flow}",
    eprint = "2201.11669",
    archivePrefix = "arXiv",
    primaryClass = "hep-ph",
    doi = "10.1016/j.cpc.2022.108565",
    journal = "Comput. Phys. Commun.",
    volume = "283",
    pages = "108565",
    year = "2023"
}

@article{Liu:2022mfb,
    author = "Liu, Zhi-Feng and Ma, Yan-Qing",
    title = "{Determining Feynman Integrals with Only Input from Linear Algebra}",
    eprint = "2201.11637",
    archivePrefix = "arXiv",
    primaryClass = "hep-ph",
    doi = "10.1103/PhysRevLett.129.222001",
    journal = "Phys. Rev. Lett.",
    volume = "129",
    number = "22",
    pages = "222001",
    year = "2022"
}

@article{Klappert:2020nbg,
    author = {Klappert, Jonas and Lange, Fabian and Maierh{\"o}fer, Philipp and Usovitsch, Johann},
    title = "{Integral reduction with Kira 2.0 and finite field methods}",
    eprint = "2008.06494",
    archivePrefix = "arXiv",
    primaryClass = "hep-ph",
    reportNumber = "TTK-20-24, P3H-20-041, FR-PHENO-2020-11, MITP/20-044",
    doi = "10.1016/j.cpc.2021.108024",
    journal = "Comput. Phys. Commun.",
    volume = "266",
    pages = "108024",
    year = "2021"
}

@article{Guan:2024byi,
    author = "Guan, Xin and Liu, Xiao and Ma, Yan-Qing and Wu, Wen-Hao",
    title = "{Blade: A package for block-triangular form improved Feynman integrals decomposition}",
    eprint = "2405.14621",
    archivePrefix = "arXiv",
    primaryClass = "hep-ph",
    doi = "10.1016/j.cpc.2025.109538",
    journal = "Comput. Phys. Commun.",
    volume = "310",
    pages = "109538",
    year = "2025"
}

@article{Smirnov:2014hma,
    author = "Smirnov, Alexander V.",
    title = "{FIRE5: A C++ implementation of Feynman Integral REduction}",
    eprint = "1408.2372",
    archivePrefix = "arXiv",
    primaryClass = "hep-ph",
    reportNumber = "SFB-CPP-14-60",
    doi = "10.1016/j.cpc.2014.11.024",
    journal = "Comput. Phys. Commun.",
    volume = "189",
    pages = "182--191",
    year = "2015"
}

@article{Binosi:2008ig,
    author = "Binosi, D. and Collins, J. and Kaufhold, C. and Theussl, L.",
    title = "{JaxoDraw: A Graphical user interface for drawing Feynman diagrams. Version 2.0 release notes}",
    eprint = "0811.4113",
    archivePrefix = "arXiv",
    primaryClass = "hep-ph",
    reportNumber = "ECT*-08-10",
    doi = "10.1016/j.cpc.2009.02.020",
    journal = "Comput. Phys. Commun.",
    volume = "180",
    pages = "1709--1715",
    year = "2009"
}

@article{Chen:2026maw,
    author = "Chen, Xiang and Guan, Xin and He, Chuan-Qi and Ma, Yan-Qing",
    title = "{Two-loop QCD corrections to $e^{+}e^{-}\rightarrow Z^{*}\rightarrow J/\psi + J/\psi$}",
    doi = "10.1103/n2sz-8b2x",
    journal = "Phys. Rev. D",
    volume = "113",
    number = "9",
    pages = "094017",
    year = "2026"
}

@article{ParticleDataGroup:2024cfk,
    author = "Navas, S. and others",
    collaboration = "Particle Data Group",
    title = "{Review of particle physics}",
    doi = "10.1103/PhysRevD.110.030001",
    journal = "Phys. Rev. D",
    volume = "110",
    number = "3",
    pages = "030001",
    year = "2024"
}

@article{Herren:2017osy,
    author = "Herren, Florian and Steinhauser, Matthias",
    title = "{Version 3 of RunDec and CRunDec}",
    eprint = "1703.03751",
    archivePrefix = "arXiv",
    primaryClass = "hep-ph",
    reportNumber = "TTP17-011",
    doi = "10.1016/j.cpc.2017.11.014",
    journal = "Comput. Phys. Commun.",
    volume = "224",
    pages = "333--345",
    year = "2018"
}

@article{Chetyrkin:2000yt,
    author = "Chetyrkin, K. G. and Kuhn, Johann H. and Steinhauser, M.",
    title = "{RunDec: A Mathematica package for running and decoupling of the strong coupling and quark masses}",
    eprint = "hep-ph/0004189",
    archivePrefix = "arXiv",
    reportNumber = "DESY-00-034, TTP-00-05",
    doi = "10.1016/S0010-4655(00)00155-7",
    journal = "Comput. Phys. Commun.",
    volume = "133",
    pages = "43--65",
    year = "2000"
}

@article{Bodwin:2007fz,
    author = "Bodwin, Geoffrey T. and Chung, Hee Sok and Kang, Daekyoung and Lee, Jungil and Yu, Chaehyun",
    title = "{Improved determination of color-singlet nonrelativistic QCD matrix elements for S-wave charmonium}",
    eprint = "0710.0994",
    archivePrefix = "arXiv",
    primaryClass = "hep-ph",
    reportNumber = "ANL-HEP-PR-07-48",
    doi = "10.1103/PhysRevD.77.094017",
    journal = "Phys. Rev. D",
    volume = "77",
    pages = "094017",
    year = "2008"
}

@article{Chung:2008km,
    author = "Chung, Hee Sok and Lee, Jungil and Yu, Chaehyun",
    title = "{Exclusive heavy quarkonium + gamma production from e+ e- annihilation into a virtual photon}",
    eprint = "0808.1625",
    archivePrefix = "arXiv",
    primaryClass = "hep-ph",
    doi = "10.1103/PhysRevD.78.074022",
    journal = "Phys. Rev. D",
    volume = "78",
    pages = "074022",
    year = "2008"
}

@article{Jacob:1959at,
    author = "Jacob, M. and Wick, G. C.",
    title = "{On the General Theory of Collisions for Particles with Spin}",
    doi = "10.1006/aphy.2000.6022",
    journal = "Annals Phys.",
    volume = "7",
    pages = "404--428",
    year = "1959"
}

@inproceedings{Haber:1994pe,
    author = "Haber, Howard E.",
    title = "{Spin formalism and applications to new physics searches}",
    booktitle = "{21st Annual SLAC Summer Institute on Particle Physics: Spin Structure in High-energy Processes (School: 26 Jul - 3 Aug, Topical Conference: 4-6 Aug) (SSI 93)}",
    eprint = "hep-ph/9405376",
    archivePrefix = "arXiv",
    reportNumber = "SCIPP-93-49, NSF-ITP-94-30",
    pages = "231--272",
    month = "4",
    year = "1994"
}

@article{Xu:2012uh,
    author = "Xu, Jia and Dong, Hai-Rong and Feng, Feng and Gao, Ying-Jia and Jia, Yu",
    title = "{Exclusive decay of {\ensuremath{\Upsilon}} into J/{\ensuremath{\psi}}+$\chi_{c0,1,2}$}",
    eprint = "1212.3591",
    archivePrefix = "arXiv",
    primaryClass = "hep-ph",
    doi = "10.1103/PhysRevD.87.094004",
    journal = "Phys. Rev. D",
    volume = "87",
    number = "9",
    pages = "094004",
    year = "2013"
}

@article{Brambilla:2017kgw,
    author = "Brambilla, Nora and Chen, Wen and Jia, Yu and Shtabovenko, Vladyslav and Vairo, Antonio",
    title = "{Relativistic corrections to exclusive $\chi_{cJ} + \gamma$ production from $e^+ e^-$ annihilation}",
    eprint = "1712.06165",
    archivePrefix = "arXiv",
    primaryClass = "hep-ph",
    reportNumber = "TUM-EFT-68-15, TUM-EFT 68/15",
    doi = "10.1103/PhysRevD.97.096001",
    journal = "Phys. Rev. D",
    volume = "97",
    number = "9",
    pages = "096001",
    year = "2018",
    note = "[Erratum: Phys.Rev.D 101, 039903 (2020)]"
}

@article{Sang:2023hjl,
    author = "Sang, Wen-Long and Yang, De-Shan and Zhang, Yu-Dong",
    title = "{Z-boson radiative decays to an S-wave quarkonium at NNLO and NLL accuracy}",
    eprint = "2302.06439",
    archivePrefix = "arXiv",
    primaryClass = "hep-ph",
    doi = "10.1103/PhysRevD.108.014021",
    journal = "Phys. Rev. D",
    volume = "108",
    number = "1",
    pages = "014021",
    year = "2023"
}

@article{Sang:2022erv,
    author = "Sang, Wen-Long and Yang, De-Shan and Zhang, Yu-Dong",
    title = "{Z boson radiative decays to a P-wave quarkonium at NNLO and LL accuracy}",
    eprint = "2208.10118",
    archivePrefix = "arXiv",
    primaryClass = "hep-ph",
    doi = "10.1103/PhysRevD.106.094023",
    journal = "Phys. Rev. D",
    volume = "106",
    number = "9",
    pages = "094023",
    year = "2022"
}

@article{Zhang:2023mky,
    author = "Zhang, Yu-Dong and Bai, Xiao-Wei and Feng, Feng and Sang, Wen-Long and Zhou, Ming-Zhen",
    title = "{Two-loop QCD corrections to C-even bottomonium exclusive decays to double J/{\ensuremath{\psi}}}",
    eprint = "2310.07453",
    archivePrefix = "arXiv",
    primaryClass = "hep-ph",
    doi = "10.1103/PhysRevD.108.114030",
    journal = "Phys. Rev. D",
    volume = "108",
    number = "11",
    pages = "114030",
    year = "2023"
}
\bibliographystyle{JHEP}
\end{document}